\documentclass{ees2023}
\usepackage{graphicx}
\usepackage{hyperref}
\usepackage[]{natbib}  
\usepackage{epstopdf}

\def\BibTeX{{\rm B\kern-.05em{\sc i\kern-.025em b}\kern-.08em
    T\kern-.1667em\lower.7ex\hbox{E}\kern-.125emX}}
\bibpunct{(}{)}{;}{a}{}{,}  %%%%%%%%%%%%%  A&A bibliography style

\begin{document}

\TitreGlobal{EES 2023}

%%-----------------------------------------------------------------
%%      the top matter
%%

\title{The Variable Sources in the Gaia archive}

\author{L. Eyer}
\address{D\'epartement d'Astronomie de l'Universit\'e de Gen\`eve, Chemin Pegasi 51, 1290 Versoix, Swizterland}
\email{laurent.eyer@unige.ch}

%\author{J.-P. Author2}\address{Institute XYZ, 1299 City, OtherLand}

%% IF Author3 has the same affiliation than Author1:
%\author{C.\,E. Author3$^1$}

%% IF Author3 has its own affiliation:
%\author{C.\,E. Author3}\address{Dept. of Chess, University of Games, 35101 Las Vegas, Monaco} 

%% IF Author3 has two affiliations, the one of Author1 and a second one:
%\author{C.\,E. Author3$^{1,}$}\address{Dept. of Chess, University of Games, 35101 Las Vegas, Monaco} 

%% Keep this line, even if the page will be settled afterwards.
\setcounter{page}{1}

%%-----------------------------------------------------------------

\maketitle

%%-----------------------------------------------------------------
%%        The abstract
%% 
%%  Warning!  within the abstract:
%%  - do not use macros. 
%%  - do not use commands like: \cite, \citet, \citep ... etc.

\begin{abstract}
At the core of the Gaia mission is a multi-epoch survey consisting of astrometric, photometric, spectrophotometric, and spectroscopic measurements.
The astrometric time series provides parallaxes and proper motions, along with information on astrometric binary systems.
The photometric time series offers a means to investigate the variability of the sources.
Due to their whole-sky, multi-epoch nature, multiple instruments, their magnitude range covering 21 magnitudes, and their remarkable photometric precision, these data allow us to describe the variability of celestial phenomena in an unprecedented manner.
For the third Gaia Data Release (DR3), the data collection spanned 34~months, with a median number of field-of-view measurements in the G band of about 44, reaching up to 270.
At publication time, DR3 delivered the largest collection of variable sources with an associated classification across the entire sky.
All these sources have their $G$, $G_{BP}$, $G_{RP}$ epoch data published and accessible in the Gaia ESA archive.
In summary, there are 10.5~million variable sources, including 9.5~million variable stars and 1~million QSOs.
Additionally, 2.5~million galaxies were identified thanks to spurious variability caused by the non-axisymmetric nature of galaxies and the way Gaia collects data.
Moreover, all the epoch data and time series of nearly 1.3~million sources in a pencil beam around the Andromeda galaxy are published, regardless of their status (constant or variable); 
This dataset is known as the Gaia Andromeda Photometric Survey.
We also introduce the citizen science project, GaiaVari to classify variable stars, the Focused Product Release delivered on October 10, 2023.
%and the Gaia BH3: the black hole detected solely with Gaia data, having time series of astrometry and radial velocitiy.
In the future, DR4 will cover 66 months, and we hope DR5 will have accumulated 10.5 years of data.
\end{abstract}

%% Insert the A&A type keywords (to appear in the ADS indexing)

%%-----------------------------------------------------------------

\section{Introduction}
%%---------------------
In astronomy, from its beginnings to now, the primary means of learning about the Universe and its constituents has been through the photons.
We can find the following text\footnote{Translated from French.}  by Evry Schatzman in his book "Astrophysique" \cite{Schatzman1963}:
\begin{quote}
\it
The fundamental basis of astrophysics lies in the radiation that reaches us from celestial objects. Whether in the visible or invisible spectrum, it is always a form of radiation that we measure.
\end{quote}

Only from the late 20th century were alternative methods used to learn about our Universe, first as exceptions to the sentence above, such as the cosmic rays \citep{Hess1912}, then with important implications, such as the neutrinos (from the Sun, \citealt{Davies1968}),  and the gravitational waves in 2015  \citep{2016PhRvL.116f1102A}.

The classical subdivision in observational astronomy about the three main observation branches are:
\begin{itemize}
	\item Astrometry: Knowledge of celestial objects obtained from their positions, motions, and shapes. Probably this discipline is as old as humanity.
	\item Photometry: Knowledge of celestial objects obtained from the measurement of their integrated light over a range of wavelengths. Historically, photometry went hand in hand with astrometry as they are the most basic perceptions of the human eye. The magnitude scale, as it is well known, is inherited from the ancient greeks (and formalized by Norman Pogson, cf.~\citealt{Pogson1856}).
	\item Spectroscopy: Knowledge of celestial objects obtained from their spectra. This technique emerged in the 19th century, with Joseph von Fraunhofer discovering dark lines in the spectra of the Sun and other stars, and Gustav Kirchhoff and Robert Bunsen correctly interpreting these dark lines as being caused by specific atomic elements (for a brief historical account see \citealt{Apprenzeller2013}).
\end{itemize}

It should be remarked that we gain knowledge not only from the source but also from what is between the source and the observer, e.g.~the interstellar medium. Observations from the ground can also allow us to determine the properties of the atmosphere and its evolution, such as in \cite{BurkiEtal1995}. 

Each of these three main observation branches benefits from multi-epoch observations. The most knowledge is obtained when these different multi-epoch domains are combined. Such merging is fruitful for finding and calibrating standard candles, identifying stellar populations, probing the invisible with microlensing events (e.g. merging the photometry and the astrometry as in \citealt{SahuEtal2022}) and determining astrophysical parameters thanks to binaries \citep{EyerEtal2015}, to pulsating stars (Baade-Wesselink method, \citealt{CarneyEtal1992}), etc.

It should be noted that on the side of photometric surveys, there has been a booming period for these past 30 years with  OGLE \citep{OGLE2003}, HAT \citep{HAT2004}, Kepler \citep{BoruckiEtal2010}, and ZTF \citep{BellmEtal2019}, to name just a few. The future is also bright with the LSST \citep{IvezicEtal2019} and PLATO \citep{RauerEtal2014}, again to name just a few.  Spectroscopy is entering into the era of systematic multi-epoch global surveys, obviously with Gaia, but also with 4MOST \citep{deJongEtal2019}, and in the future with WST \citep{MainieriEtal2024}.

Now, if we turn to the Gaia cornerstone mission, the "tour de force" of Gaia is to assemble multi-epoch data for the entire sky on these three main observation branches: astrometry, photometry, and spectroscopy.

In this short text, we will mostly restrict the presentation to what has been put in the Gaia archive for the variable sources.

\section{The Gaia mission: An introduction}
%%---------------------
Let us review key aspects of the Gaia mission: Gaia is a cornerstone mission of the European Space Agency science program \citep{PrustiEtal2016}, tasked with performing a systematic survey of all objects brighter than G=20.7, attaining a survey of more than 2 billion celestial sources.
The measurements collected by Gaia consist of position (astrometry), brightness, and colours (photometry and spectrophotometry), along with radial velocity (spectroscopy). It should be noted that the survey for radial velocities is restricted to brighter sources (about $G=15$).
Gaia was launched on December 19, 2013 by a Soyuz rocket. The CCD camera aboard Gaia is the largest ever deployed in space, reaching nearly one billion pixels.
Over its ten-year mission, each source, on average, will be observed 140 times in each of the 9 CCDs\footnote{for each of the 7 raws at the exception of raw 5, for which the number is 8.}  in the G band, as well as in each of the BP and RP CCDs. The number of measurements for the radial velocity spectrometer instrument will be comparatively fewer, averaging approximately 80 times.
The results of the Gaia mission, that the Data Processing and Analysis Consortium processes, are made accessible to the public through Data Releases. DR4 and DR5 are anticipated to be delivered in 2026 and no later than 2032, respectively.

We will focus on this article on the DR3 results, which compiles 34 months of observations and nearly one trillion CCD measurements over the whole celestial sphere.

Several factors contribute to Gaia's exceptional status as a mission:
\begin{itemize}
    \item Unprecedented Astrometry: Gaia provides unparalleled astrometric data, i.e. positions, parallaxes, and proper motions for more than a billion stars over the whole celestial sphere. The astrometric precision is available on the Gaia webpage (\url{https://www.cosmos.esa.int/web/gaia}\footnote{Note: This page has undergone changes in URL and error estimates following the better understanding of the performance.}): For Gaia DR4, that is, for the nominal mission, the parallax uncertainty is 22~$\mu$as at magnitude 15. This precision can be compared to the 20-25~$\mu$as estimates from pre-launch assessments in 2006 for a G2 star at magnitude 15 (see  e.g. \citealt{Eyer2006}). It's important to note that the errors in position and proper motion need to be multiplied by 0.8 and 0.5, respectively.
The astrometry is also able to detect astrometric binaries, exoplanets, microlensing effects, etc\ldots
    \item Three Instruments on a Single Platform: astrometry, [spectro-]photometry, spectroscopy with the derivation of radial velocities. This results in a unprecedented uniformity of data across the entire celestial sphere.
    \item Multi-Epoch Measurements: The mission conducts multi-epoch measurements of the entire sky, providing a comprehensive view of variable celestial objects, see also Section~\ref{timesampling}.
    \item 10-Year Time Baseline: With a 10-year mission duration (if all goes as planned), Gaia offers high frequency precision, particularly beneficial for periodic objects.
    \item Quasi-Simultaneous Measurements: Gaia achieves quasi-simultaneous measurements. Other multi-band photometric surveys do not have this simultaneity and this complicates somewhat the analysis. For more details, see Section~\ref{timesampling}.
    \item Time-Domain Selection Function: Because of its predefined scanning law, Gaia allows us to determine the time-domain selection function, enabling the understanding of observational biases. 
    \item Extensive Number of Measurements: Gaia's data volume is significant, with nearly 1 trillion CCD measurements for DR3 (i.e. 34 months).
    \item High Dynamical Range: Gaia covers a high dynamical range, capturing data from the brightest sources (G $\approx 1.7$ in DR3) to magnitudes as faint as G $\approx 21/22$.
    \item Space-based Operation: Gaia's position in space provides stability and allows us to access the entire celestial sphere from a single platform, which is not achievable by any single ground-based optical telescope. However, operating in space presents its own challenges, such as the high cost of space missions\footnote{As Prof. Bohdan Paczynski humorously noted, "One dollar in space is worth less than one dollar on Earth".}, the impossibility of repairs, micrometeoroid hits, and solar flares/the radiation damages. Indeed Gaia's location at L2, 1.5 million kilometers from Earth in the direction opposite to the Sun, leaves it unshielded by Earth's magnetic field.
    \item Cyclic Improvements: The Gaia consortium implements cyclic improvements, performing systematic data analyses, with not only more data at each cycle, but with improved calibrations, and enhanced outlier detection.
\end{itemize}     
     
Each of these elements makes Gaia extremely fruitful for variable and binary stars. These are among the objects that benefit the most from Gaia.
Due to the semi-regular Gaia sampling, strictly periodic objects can be studied in detail despite the gaps in the scanning law.  Also transient sources are and will be classified in the data releases \citep{EyerEtal2023,RimoldiniEtal2023} and are also detected in real time by the Gaia Science Alert System \citep{HodgkinEtal2021}. Indeed, since the beginning of its science operations, Gaia has continuously (with few interruptions) provided alerts, identifying potentially time-sensitive events. If not addressed promptly, these events could lead to significant scientific loss. To date (June 2024), the Science Alerts team has released over 25,000 of these alerts. About 27\% of the alerts are classified, with approximately 60\% of these classified alerts identified as supernovae.

\section{The Gaia time sampling}
\label{timesampling}
The time sampling\footnote{referred to as the cadence in LSST/Vera C. Rubin observatory.} of Gaia is very particular and semi-regular. It is optimized to achieve astrometric precision as uniform as possible across the entire celestial sphere. This sampling is known as the Nominal Scanning Law (NSL): Gaia continuously sweeps the sky, with the spacecraft's rotation axis precessing on a cone with a 45-degree opening angle to the Sun. A full precession cycle on the cone is completed every 63 days. The spacecraft rotates on its axis every 6 hours, and the 106.5-degree angle between Gaia's two fields-of-view (preceding and following fields-of-view) results in a time interval of 1 hour and 46 minutes between them  (the exemplary angle of 253.5-degree between the following and preceding fields-of-view takes thus 4 hours 14 minutes).

\begin{figure}[htbp]
    \centering
    \includegraphics[width=0.8\linewidth]{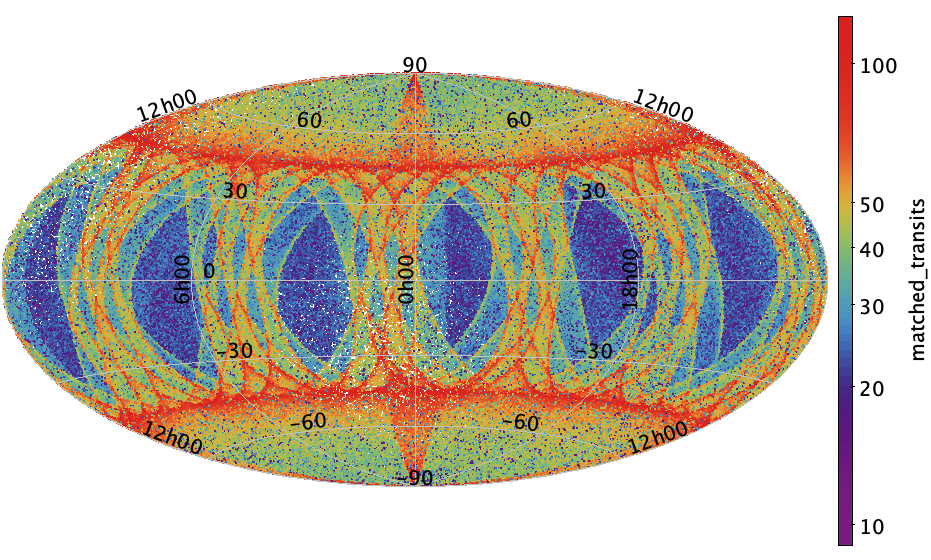}
    \vspace{1em} % optional space between the images
    \includegraphics[width=0.8\linewidth]{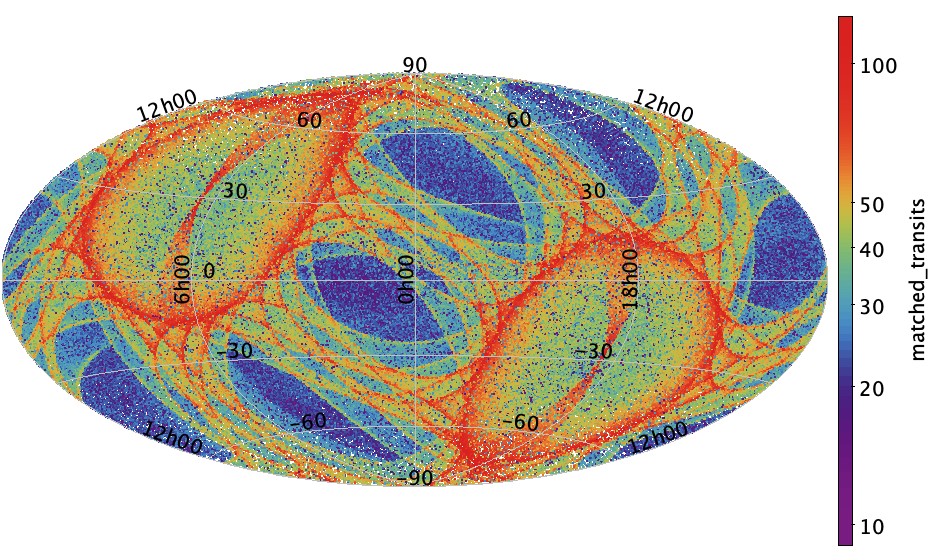}
    \caption{Number of fields-of-view data per source in ecliptic coordinates (top) and Galactic coordinates (bottom) for DR3 in Aitoff projections. We see the complex sampling which is clearly not completed yet, with blue regions, i.e. low number of transtis, around the ecliptic plane. We see that one of these low number of measurements is in the direction of the lower bulge at positive Galactic longitudes.}
    \label{fig:skynumfov}
\end{figure}

Figure~\ref{fig:skynumfov} shows the sky plot for the number of observations (the field-of-view transit). We took the field \texttt{matched\_transits}, i.e. the total number of field–of–view transits matched to this source. This number is higher or equal to the \texttt{num\_selected\_g\_fov} of \texttt{gaiadr3.vari\_summary} table, because it does not take into account the number of observations that are removed.

\begin{figure}[htbp]
    \centering
    \includegraphics[width=0.85\linewidth]{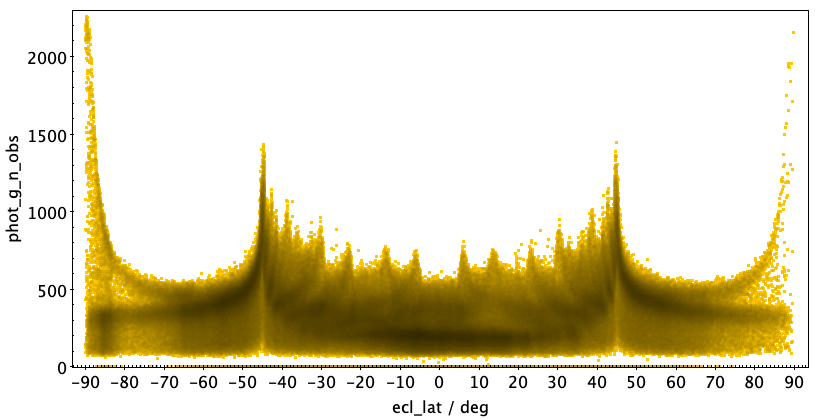} % Replace 'filename' with the actual filename of your image
    \caption{Number of CCD observations per source as a function of ecliptic latitude ($\beta$) for DR3. A large number of observations are visible at $\pm 45^\circ$ due to the Nominal Scanning Law. Additionally, stars at the ecliptic poles (both ends of the plot) have an exceptionally high number of observations, exceeding 2000, due to the Ecliptic Pole Scanning Law, which was in effect for the first 28 days of the mission.}
    \label{fig:ccdobsecllat}
\end{figure}

The second plot, Fig.~\ref{fig:ccdobsecllat}, presents the number of CCD observations as a function of ecliptic latitude. On a field crossing the CCD observations are separated by 4.85 seconds. The Annex A of \cite{EyerEtal2017} provides detailed properties of the scanning law for 5 years, including various representations, histograms of the lags, and spectral windows for different regions of the sky.

Because the scanning law is well-defined, it is possible to study the selection function. \cite{EyerMignard2005} conducted a study on periodic signals based on an earlier design of the spacecraft and a 5-year scanning law. They concluded that the ability to detect the correct period of a periodic signal strongly depends on the ecliptic latitude. For DR3, significant differences in detection exist even at fixed ecliptic latitudes due to the scanning law creating regions with a low number of observations; see Fig.~\ref{fig:skynumfov}.

There are two exceptions to the Nominal Scanning Law:
\begin{enumerate}
 \item At the beginning of the mission, a 28-day period was planned to assist with photometric calibrations. During this time, the same stars and regions of the sky were scanned nearly continuously. The axis of rotation of the spacecraft was kept in the ecliptic plane (at a 45-degree angle to the Sun) without precession, allowing the ecliptic poles to be scanned regularly during these 28 days, with repeated sequences of intervals 1h46m, 4h14m, 1h46m, etc. As the spacecraft orbits around the Sun, stars at lower ecliptic latitudes (in absolute value) are observed for shorter sequences.
 \item The second exception occurred when, for 12 months\footnote{6 months will be included in DR4, the rest in DR5.}, a modified NSL was introduced to improve the astrometric solution. As a positive side effect it reduces aliasing peaks. In this mode, the precession of the spacecraft's rotation axis was reversed. However, the spacecraft's rotation direction remained unchanged due to the constraints of the Time Delay Integration (TDI) mode. Note that the reverse motion of the rotation axis on the cone increased the consumption of cold gas.
\end{enumerate}

%At the level of field-of-view crossing, the measurements in the three instruments are nearly simultaneous. The field crossing The separation between them:
%at the level of G band magnitude is made by 9 measurements of CCDs.

Although the scanning law determines which stars are observed, understanding the temporary absence of an observed source is challenging. Several factors could explain a missing observation: the star might have truly disappeared (the most interesting case) by becoming very faint due to an eclipse or occultation, it could be just below the detection threshold of the less precise sky mapper and thus not observed, the star can be perturbed by the other field-of-view (e.g. bright star/extended object) which is superposed on the same focal plane, the data might have been deleted due to excessive data volume onboard, or technical problems might have occurred.

%\begin{figure}[htbp]
%    \centering
%    \includegraphics[width=0.85\linewidth]{nobs_fov_sky.png} % Replace 'filename' with the actual filename of your image
%    \caption{Number of field of view data per source in ecliptic coordinates (Aitoff projection) for DR3. We see the complex sampling which is clearly not completed yet, with green regions around the ecliptic plane.}
%    \label{fig:skynumfov}
%\end{figure}

\section{The Gaia G-band photometric precision}

\begin{figure}[htbp]
    \centering
    \begin{minipage}{0.49\linewidth}
        \centering
        \includegraphics[width=\linewidth]{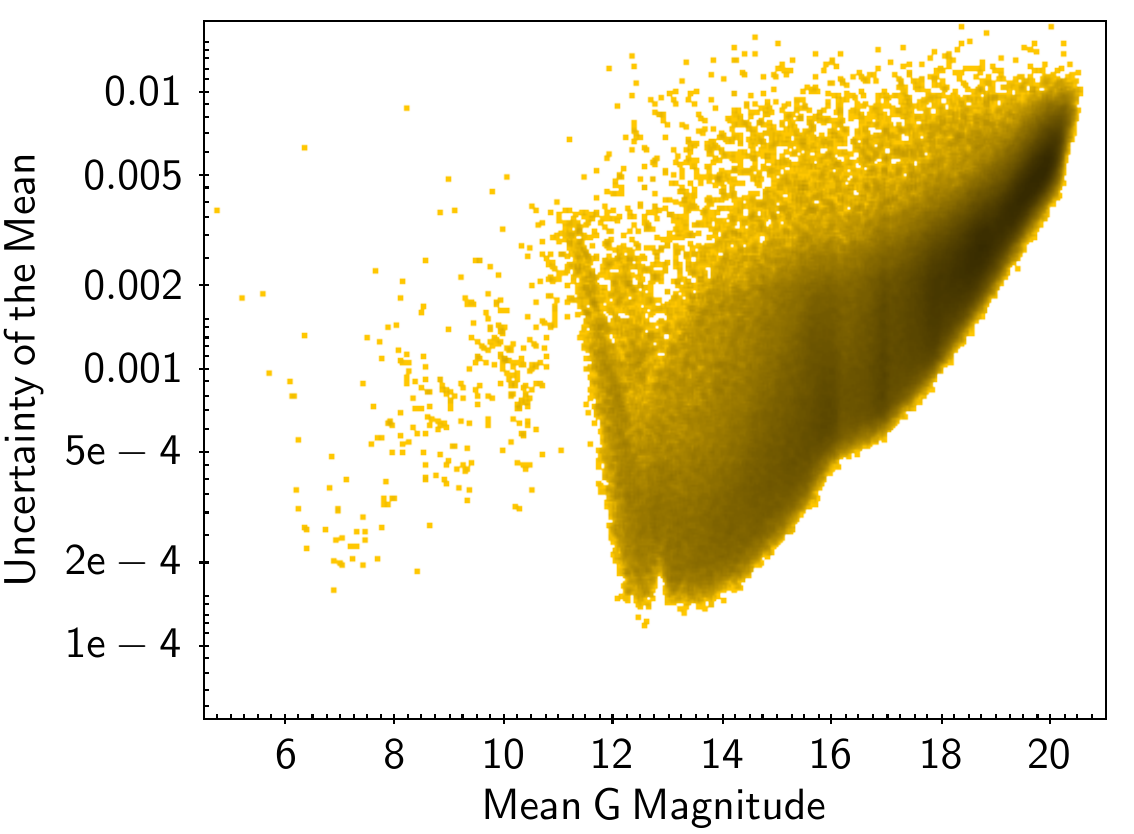}
    \end{minipage}
    \hfill
    \begin{minipage}{0.49\linewidth}
        \centering
        \includegraphics[width=\linewidth]{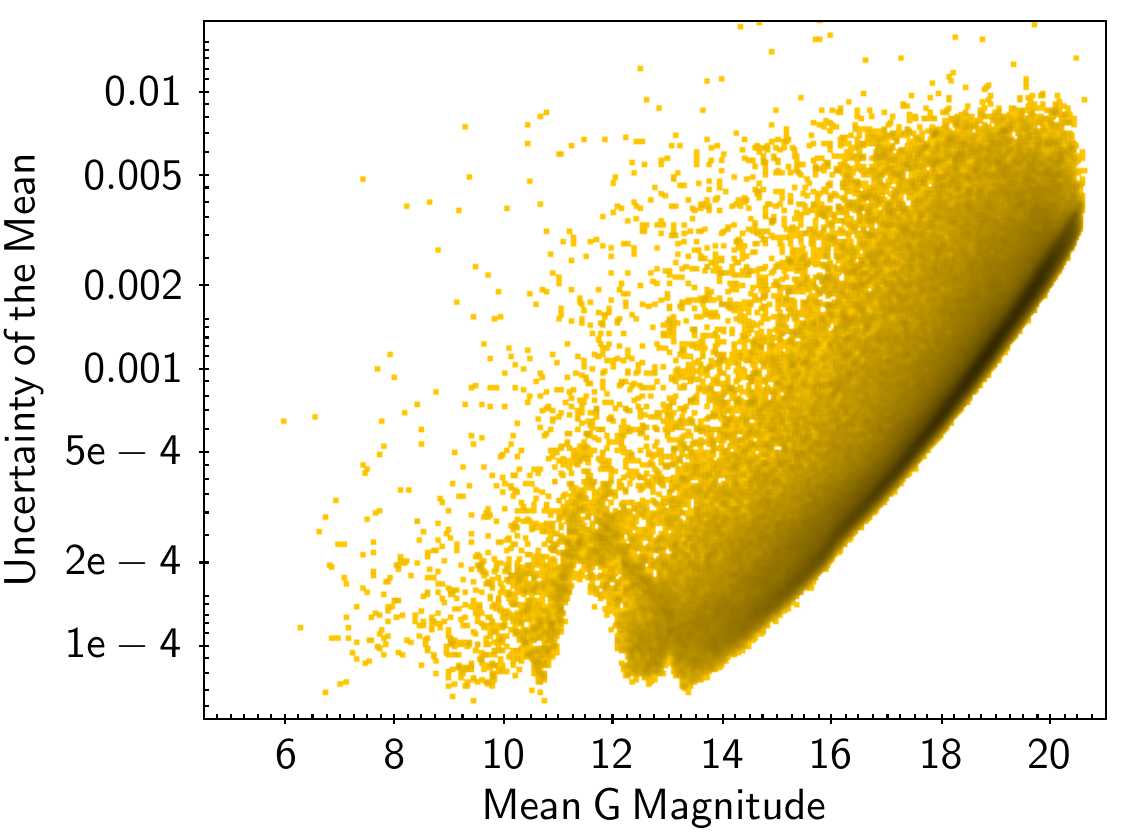}
    \end{minipage}
    \caption{Uncertainty on the mean versus the mean G magnitude. Two plots are displayed side by side, for the comparison of DR1 (left) and DR3 (right) photometry. We see that not only has the noise gone down thanks to the square root of the number of measurements but also that the calibration has significantly improved. The various changes in windowing schemes or gates are better taken into account in DR3 than in DR1.}
    \label{fig:Gbandphotomprecision}
\end{figure}

Fig.~\ref{fig:Gbandphotomprecision}  shows the uncertainty in the mean G-band photometry as a function of the G magnitude.
There are two plots: the results from DR1 and DR3.
DR2 is not displayed to avoid too many figures and it falls between the two.
The plot is based on a selection of stars with absolute ecliptic latitudes below 60 degrees to avoid the perturbation of ecliptic pole scanning.
The uncertainty is estimated from the flux standard deviation divided by the square root of the number of measurements (after some clipping).
Naturally, this uncertainty decreases as the number of measurements increases.
In case of variable sources, the standard deviation is obviously larger.
Although more measurements may reveal more instrumental calibration features, this is not the case here.
On the contrary, some features visible in DR1 are less prominent or have disappeared in DR3 demonstrating the improvements of the calibrations.
It should be noted that in the Gaia archive, uncertainties are provided in flux rather than in magnitude.
Therefore, these uncertainties need to be converted into magnitude uncertainties for Fig.~\ref{fig:Gbandphotomprecision}. We used the formula $\mbox{Uncertainty(magnitude)}= 1.086 \times \mbox{Uncertainty(flux)}/\mbox{flux}$, where the term 1.086 is coming from $2.5/\ln(10)$.

We will not enter into the debate on using flux versus magnitudes to evaluate the brightness of a celestial source here, opening Pandora's box.
The main point is to avoid being too dogmatic about these issues and to use the approach that makes the most sense for the problem at hand.
As a side note, negative fluxes were unfortunately excluded in the mean calculation process \citep{RielloEtal2021, EvansEtal2023}.
Negative fluxes can occur due to statistical variations or overestimation of the background for example.
This exclusion introduces a bias in the magnitude estimation at the faint end: the sources are brighter than they are in reality.
Starting from DR4, negative fluxes will be included.

%\begin{figure}[htbp]
%    \centering
%    \includegraphics[width=0.85\linewidth]{DR1_phot.png,DR3_phot.png}
    % Replace 'filename' with the actual filename of your image

%    \caption{Uncertainty on the mean versus the mean G magnitude.
%    Two plots are superposed,  for the comparison of DR1 (front) and DR3 photometry (back). 
%    We see that not only has the noise gone down thanks to the square root of the number of measurements but also that the calibration has significantly improved.
%    The various changes in windowing schemes or gates are better taken into account in DR3.
%    }
%    \label{fig:Gbandphotomprecision}
%\end{figure}

\section{The Gaia data releases and the variability}
%My introduction.
%Gaia \cite{2016A&A...595A...1G} DR3 \cite{2023A&A...674A...1G}

The data release process is iterative and advances with each step in both quantity and quality, this is particularly evident for the variability processing and analysis.
So let us have a quick look at the DR1 and DR2 data releases.
A global pipeline was set up for the variability processing and analysis of Gaia data. Initially, it was applied to the cycle 1 data,  the details of this processing can be found in \cite{EyerEtal2017}.
The DR1 publication also covers the data properties of cycle 1, along with certain aspects of the 5-year sampling obtained through simulations.
Based on 14 months of data, the DR1 variable star catalogue showcased a selection of 3,194 variable stars. 
This data release consisted of two types of variability, namely Cepheids and RR Lyrae stars in the direction of the Large Magellanic Cloud \citep{ClementiniEtal2016}.
The Large Magellanic Cloud, situated near the South Ecliptic pole, benefited from the regular sampling during the 28 days of the Ecliptic Pole Scanning Law.
In addition, a performance verification paper \citep{ClementiniEtal2017} presented the TGAS parallaxes for Cepheids and RR Lyrae stars, demonstrating that already the improvements with respect to Hipparcos were impressive.

The DR2 based on 22 months of operation data reached 550,737 variable stars \citep{HollEtal2018}.
The DR2 contained the following variability types: RR Lyrae stars, Cepheids, long period variables, $\delta$ Scuti/SX Phoenicis stars, stars with rotation modulation and stars with short time scale variability.
A performance verification paper \citep{EyerEtal2019} presented the properties of variable sources across the Hertzsprung-Russel Diagram. As a remarkable result, we can follow the path of stars, their motion in the Hertzsprung-Russel Diagram as their magnitude and colour change.

More generally, information and details for all data releases can be found in various articles (many are listed in this section for details on variability processing and analysis) or
in the Gaia documentation associated with the data releases on the archive website, \url{https://gea.esac.esa.int/archive/documentation/GDR3/} for DR3. This document, available online and also in PDF format, contains more than 1,500 pages.

It is also important to consult the "known issues" page. For each data release, there is a section on the Data tab of \url{https://www.cosmos.esa.int/web/gaia/data}. Under these sections, you can find the known issues for each data release. For example, the known issues for DR3 can be found at the following address: \url{https://www.cosmos.esa.int/web/gaia/dr3-known-issues}.

A note on the archive names:
you have in the ESA Gaia archive (\url{https://gea.esac.esa.int/archive/}) the different data sets associated with the different data releases,
\texttt{gaiadr1.table\_name},
\texttt{gaiadr2.table\_name},
\texttt{gaiaedr3.table\_name},
\texttt{gaiadr3.table\_name},
\texttt{gaiafpr.table\_name} which relates the Gaia Data Release 1 (DR1), Gaia Data Release 2 (DR2), Gaia Early Data Release 3 (EDR3), Gaia Data Release 3 (DR3), Focused Product Release (FPR) respectively.
For a given source, the source\_id can change. A last remark on the nomenclature related to variability: for the tables names associated to the variability processing and analysis, starting from the second data release, these table names start with "\texttt{vari\_table\_name}".

In the following sections, we will focus more on the DR3 as it has the most numerous outputs and diversity in variability types.

\subsection{The DR3 results}
For the third data release of Gaia, the variability analysis is described in \cite{EyerEtal2023}.
In summary, as of the data release in June 2021, Gaia provided the largest catalogue ever of classified variable sources across the entire sky.
In DR3, we have 10.5~million variable sources, comprising 9.5~million variable stars and 1~million AGN.
Additionally, thanks to a spurious perturbation in the photometry, we identified 2.5~million galaxies.

\begin{figure}[htbp]
    \centering
    \includegraphics[width=0.40\linewidth]{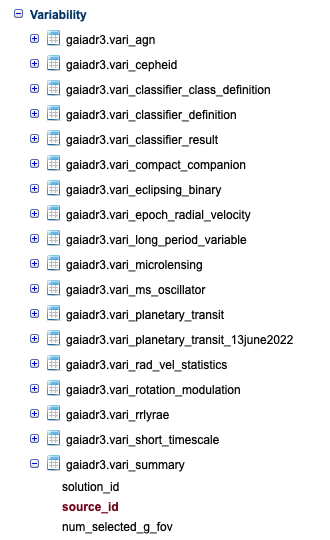} % Replace 'filename' with the actual filename of your image
    \caption{Tables available related to the variability processing and analysis in the Gaia archive. We opened the  \texttt{vari\_summary} table with just the first three lines.}
    \label{fig:tables}
\end{figure}

A general overview of the DR3 result, as mentioned above, can be found in \cite{EyerEtal2023}.
In the Gaia archive \url{https://gea.esac.esa.int/archive/}, the different associated tables related to the variability processing and analysis can be seen in Fig.~\ref{fig:tables}.

A vast and systematic compilation of the literature was done in \cite{GavrasEtal2023}; it combines the sources of 152 catalogues crossmatched with the Gaia data and contains 4.9~million variable sources.
The catalogue is available at the CDS (Centre de Donn\'ees astronomiques de Strasbourg, \url{https://cds.unistra.fr},  \cite{GavrasEtalCat2022}).

Part of this compilation was used as a training set for the supervised classification \citep{RimoldiniEtal2023}.
For the classification, the methods employed are Random Forest and eXtreme Gradient Boosting models in a mixture of multi-class and binary classifiers.
There are 24 variability (grouped) classes. Some of these are combined types. Therefore, when counting the subclasses, we reach a total of 35 variability types. The variability classes list can be found in Table~\ref{tab:vartypes}. 
\begin{table}
    \centering
    \begin{tabular}{|p{8cm} |p{8cm}  |}
        \hline
        \textbf{Variability types} & \textbf{Variability types} \\
        \hline
(1) ACV|CP|MCP|ROAM|ROAP|SXARI: $\alpha$2 CVn, (Magnetic) Chemical Peculiar, Rapidly Oscillating Am/Ap\footnote{here probably the small oscillations are not detected.} SX Arietis stars & (13) MICROLENSING: Star with microlensing event \\
(2) ACYG:  Alpha Cygni-type variable & (14) RCB: R Coronae Borealis stars \\
(3) AGN: Active Galactic Nuclei (including Quasars) &  (15) RR: RR Lyrae stars \\
(4) BCEP: Beta Cephei type variable & (16) RS: RS Canum Venaticorum type variable \\
(5) BE|GCAS|SDOR|WR: Subset of eruptive variable types: B-type emission line star, Gamma Cassiopeiae, S Doradus, and Wolf-Rayet & (17) S: Set of stars with short timescale variability \\
(6) CEP: Cepheid variable types: delta Cepheid, anomalous Cepheid, and type-II Cepheid & (18) SDB: Subdwarf B stars \\ 
(7) CV: Cataclysmic variable & (19) SPB: Slowly Pulsating B-star variable \\
(8) DSCT|GDOR|SXPHE: delta Scuti, gamma Doradus, and SX Phoenicis & (20)  SN: Supernovae \\
(9) ECL:  Eclipsing Binaries of types: Beta Persei (Algol), Beta Lyrae, and W Ursae Majoris & (21) SOLAR\_LIKE: Stars with solar-like variability due to flares, spots, and rotational modulation \\
(10) ELL:  Ellipsoidal variable & (22) SYST: Symbiotic variable star \\
(11) EP: exoplanetary transits & (23) WD: Variable White Dwarf of types: ZZ Ceti (DAV, ZZA), V777 Her (DBV, ZZB), and GW Vir (DOV, ZZO) \\
(12) LPV: Long Period Variable stars of types: omicron Ceti (Mira), OGLE Small Amplitude Red Giants, and semiregular & (24) YSO: Young Stellar Object \\   
\hline
    \end{tabular}
    \caption{The 24 variability types considered in DR3. Please note that in some cases, a denomination encompasses several different variability types.}
    \label{tab:vartypes}
\end{table}

The table that contains the variability type class from the Machine Learning is \texttt{gaiadr3.vari\_classifier\_result}. 

A validation of the classification for the young stellar object class  was done by \cite{MartonEtal2023}.

After the classification, Specific Object Studies are done for various topics. They act as a validation of the classification, and also these activities compute further parameters that describe the class.
It is important to note that there are two sources for a given variability type:
(1) The Machine Learning classification and (2) the classification of Specific Object Studies.
There may be differences between these two classifications.

\begin{figure}[htbp]
    \centering
    \includegraphics[width=0.85\linewidth]{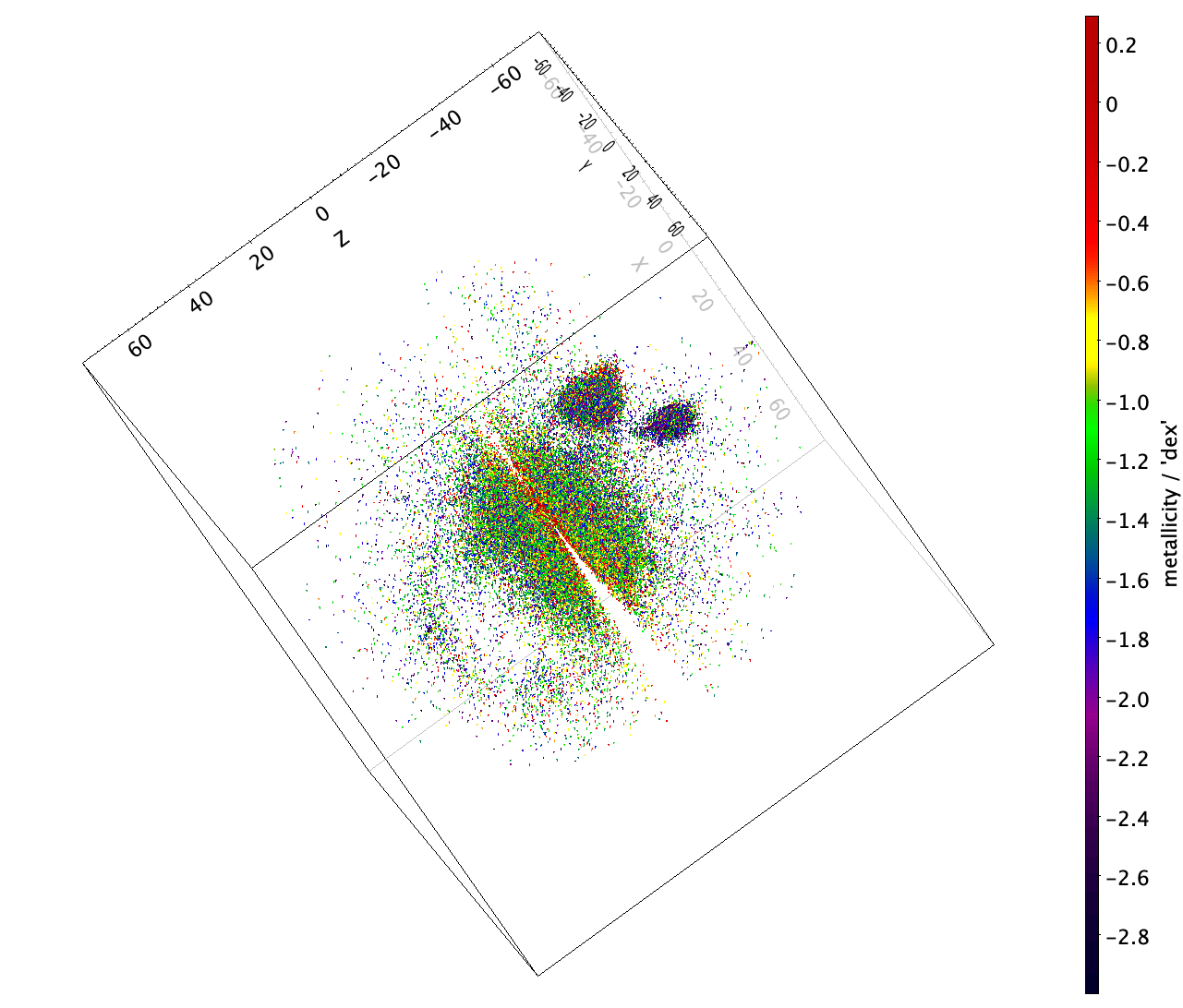} % Replace 'filename' with the actual filename of your image
    \caption{Distribution of RR Lyrae stars in a 3D projection using the distance modulus. The LMC and SMC can be seen behind the Milky Way RR Lyrae stars. The metallicity near the Galactic plane is higher, lower in the halo, then lower in the LMC and then even lower in the SMC, as expected. We also see the Sagittarius stream. We can guess the presence of other structures in the halo.}
    \label{fig:rrabmetallicity}
\end{figure}

The Specific Object Studies covers the following topics (listed in the alphabetical order, as in the Gaia archive). Here we just give the number of sources and also some selected highlights: 
\begin{itemize}
\item {\bf AGN:}  \texttt{gaiadr3.vari\_agn} \cite{CarnereroEtal2023}. There are 872,228 sources in this table, this number is obtained with the ADQL query (just shown for this example):
\begin{verbatim}
SELECT COUNT(*)
FROM gaiadr3.vari_agn
\end{verbatim}
We can compare  this number from the number of AGN from classification, we have 1,035,207 sources, see following ADQL command:
\begin{verbatim}
SELECT COUNT(*)
FROM gaiadr3.vari_classifier_result
WHERE best_class_name = 'AGN'
\end{verbatim}
A surprising result is that even with the gaps of the Gaia time series, the time delay for a source, the lens system DESJ0501-4118,  was derived.
\item {\bf Cepheids:} \texttt{ gaiadr3.vari\_cepheid} \cite{RipepiEtal2023}. There are 15,021 sources. A Fourier series model  of the light curve is fitted and the parameters are published in this table. The estimation of the mean should be taken for this table and not from \texttt{gaia\_source} table. A subclassification is also given. Radial velocities time series are published for 798 sources.  It is the largest catalogue of Cepheids having radial velocities.
 \item {\bf Compact companion}: \texttt{gaiadr3.vari\_compact\_companion} \cite{GomelEtal2023}. There are 6,306 sources. If the variability is caused by ellipsoidal deformation, a mass ratio can be determined from the Fourier parameters of the light curve. This mass ratio helps identify potential candidates for white dwarfs, neutron stars, and black holes. Other methods to detect black holes are based on astrometry and radial velocities \citep{El-BadryEtal2023,ChakrabartiEtal2023, El-BadryEtal2023, PanuzzoEtal2024}.
 
 \item {\bf Eclipsing binaries:} \texttt{gaiadr3.vari\_eclipsing\_binary} \cite{MowlaviEtal2023}. There are 2,184,477 sources. The largest catalogue of eclipsing binaries ever published on the whole sky. In this table, we adopt a geometrical modelling of the light curve and extract some relevant statistics. A sub-sample is then studied with a physical modelling, see \cite{PourbaixEtal2022}. The results are published in the non-single star table \texttt{gaiadr3.nss\_two\_body\_orbit}. It contains 86,918 systems with the flag nss\_solution\_type EclipsingBinary and  155 for EclipsingSpectro.
 \item {\bf Long Period Variables:} \texttt{gaiadr3.vari\_long\_period\_variable} \cite{LebzelterEtal2023}. There are 1,720,588 sources. Thanks to the RP spectrophotometry, 546,468 stars were classified as carbon rich candidates.
 \item {\bf Microlensing:}  \texttt{gaiadr3.vari\_microlensing} \cite{WyrzykowskiEtal2023}. This is the first whole sky microlensing event list, which contains 363 sources.
 \item {\bf Main Sequence Oscillators:} \texttt{gaiadr3.vari\_ms\_oscillator} \cite{DeRidderEtal2023}. There are 54,476 sources. See Section~\ref{Sect:PVP}.
 \item {\bf Planetary transits:} \texttt{gaiadr3.vari\_planetary\_transit} \cite{PanahiEtal2022a}. The table contains 214 sources. Please note that the table \texttt{gaiadr3.vari\_planetary\_transit\_13june2022} is incorrect. For details, please see Section~\ref{Sect:exoplanet}.
 \item {\bf Rotation modulation:} \texttt{gaiadr3.vari\_rotation\_modulation} \cite{DistefanoEtal2023}. There are 474,026 sources. The Period Amplitude diagram shows different regimes.\\
\item {\bf RR Lyrae stars:}  \texttt{gaiadr3.vari\_rrlyrae} is taking care of the RR Lyrae variables and is described in \cite{ClementiniEtal2023}. There are 271,779 sources. RR Lyrae stars are well known standard candles. There are several subtypes of them: the Bailey's ab and c types corresponding to fundamental mode pulsation and first overtone, respectively. The ab type light curves have a very recognisable shape with a sharp rise and slower declined with a period typically of half a day, the c types have a more sinusoidal light curve. An example of the quite astonishing results is the metallicity map derived from the RR Lyrae light curves in Figure~\ref{fig:rrabmetallicity}. There are 1,100 sources with radial velocities time series which are in the table of RR Lyrae stars.
\item {\bf Short Time scales:}  \texttt{gaiadr3.vari\_short\_timescale} is described in \cite{EyerEtal2023}. There are 471,679 sources.  This procedure did not go through removal of sources and reveals also instrumental and/or calibration artefacts.
\end{itemize}
For the completeness of the archive description, three additional tables are listed under the variability folder: one is a summary of the variability analysis, and the other two are associated with the radial velocities of Cepheids and RR Lyrae stars:
\begin{itemize}
\item  \texttt{gaiadr3.vari\_summary} table provides a global overview. It helps determine which table contains a given source \citep{EyerEtal2023}, and provides statistical attributes of the time series.\\
\item  \texttt{gaiadr3.vari\_epoch\_radial\_velocity}: the table contains the time series of the radial velocities.\\
\item  \texttt{gaiadr3.vari\_rad\_vel\_statistics}: the table contains statistics of the radial velocities.\\
\end{itemize}

\subsubsection{The Performance Verification Paper on variability}
\label{Sect:PVP}
A performance verification paper (PVP) authored by \cite{DeRidderEtal2023} investigate the pulsation properties in main sequence OBAF-type stars.
Among its noteworthy findings are: (1) the establishment of a period-luminosity relation for $\delta$\,Scuti stars. (2) It is further shown that stellar rotation seems to reduce the amplitude of variability.
(3) Discrepancies arise when comparing the observed locations of these variables with the predictions of theoretical models.
Such inconsistencies may be attributed to inaccuracies in determining astrophysical parameters using Gaia instruments, and/or it is conceivable that rotational effects play a significant role, potentially altering the instability regions within the Hertzsprung-Russell diagram.
(4) A quite remarkable result (see Appendix B of the work by \cite{DeRidderEtal2023}) which demonstrates Gaia performance: with as few as 40 measurements, Gaia detects two accurate frequencies in certain cases when compared with the reference observations from the Kepler mission. As Gaia extends its measurement duration to a decade, the precision of frequency determination, which scales inversely with the observation time basis, is anticipated to reach exceptional levels of accuracy.

There is also a publication on the spurious effect of spurious photometric variability introduced by the angle by which the satellite scans a (non axi-symmetric) source \citep{HollEtal2023}.

\subsubsection{The Gaia Andromeda Photometric Survey}
To offer a teaser of DR4 and DR5, which will include epoch photometry/time series for {\bf all sources}, we provide in the Gaia archive a pencil beam with a 5.5-degree opening angle centred toward the Andromeda galaxy, ensuring photometry availability for all sources within this region \citep{EvansEtal2023}.
This selection was made to provide a sufficiently broad sample of celestial sources reaching nearly 1.3~million sources; in addition to the Andromeda galaxy stars, the foreground stars, with significant parallaxes, allow us to provide a well-populated Hertzsprung-Russell Diagram. 
This dataset contains all the $G$, $G_{BP}$, $G_{RP}$ time series in the Gaia archive.
The \cite{EvansEtal2023} article also presents a renormalisation of the photometric uncertainties (as these uncertainties are usually underestimated).
We also show a way to combine the three photometric bands with a Principal Component Analysis method to enhance the  detection and the characterisation of the variable sources, an approach similar to what was done with SDSS data \citep{SuvegesEtal2012}.

\subsubsection{The exoplanets from Gaia with the transit method}
\label{Sect:exoplanet}
Although the time span of DR3 was limited to 34 months, the quality of the photometry is sufficient to discover exoplanets by the transit method \citep{PanahiEtal2022a}.  Currently, there are two exoplanets detected by the transit method and confirmed by radial velocity measurements, Gaia-1b and Gaia-2b.
The list of candidates is provided in the Gaia table  \texttt{vari\_planetary\_transit} and contains 214 sources: 173 systems are known, and 41 are new (among them the two above already cited); the list of these 41 candidates can be found in \cite{EyerEtal2023}.
There is further work with the exoplanets with the transit methods, there has been a special agreement with the TESS mission \citep{RickerEtal2015} to verify the TESS candidates by Gaia.
Again, with only 34 months, we are able (1)\,to show that about 5\% of the TESS candidates are background eclipsing binaries and also (2)\,to confirm about 5\% of the cases where the transit can be identified \citep{PanahiEtal2022b}.

As a side note, it is noteworthy to mention that: (1)\,astrometric exoplanet candidates were proposed by \cite{HollEtal2023a}.
At the time of writing, one of these candidates, named Gaia-3b, was confirmed thanks to ground-based radial velocity measurements \citep{SozzezziEtal2023}.
%Gaia-3b marks a significant milestone as it is the first exoplanet unambiguously detected thanks to astrometric measurements and confirmed via radial velocities.
(2)\,Exoplanets have also been discovered through microlensing. Gaia21dnc (Ban et al., in Prep) and Gaia22dkv \citep{WuEtal2023} are the first two exoplanets detected through follow-ups of microlensing events alerted by the Gaia Science alert system \citep{HodgkinEtal2021}.

%\subsection{The metallicity sky map from RR Lyrae stars}

%RR Lyrae stars are well known standard candles. There are several subtypes of them: the Bailey's ab and c types corresponding to fundamental mode pulsation and first overtone, respectively. The ab type light curve have a very recognisable shape with a sharp rise and slower declined with a period typically of half a day, the c types have a more sinusoidal light curve.

%As an example of quite astonishing results are the metallicty map derived from the RR Lyrae light curves.

\subsection{The focused product release: Time domain radial velocities for Long Period Variable stars}
%%--------------------
After the third data release, on October 10, 2023, there was a focused product release by the Gaia consortium, consisting of several products: 
\begin{itemize}
  \item Astrometry and photometry from engineering images taken in the $\omega$ Centauri region.
  \item The first results of quasars' environment analysis for gravitational lenses search.
  \item Extended radial velocity epoch data for Long Period Variables.
  \item Diffuse Interstellar Bands from aggregated RVS spectra.
  \item Updated astrometry for Solar System objects.
\end{itemize}

Here we present the Long Period Variables (LPV) \citep{TrabucchiEtal2023}.
The results of this focused product release are based on the DR3 processing both in photometry and radial velocities.
A reminder of DR3  on Long Period Variables: It contains 1.7~million LPV.
A selection was performed of stars fainter than $G=6$ and brighter than  $G=14$. 
The total number of sources is 9,614 stars with radial velocities times series.
Though this number may look small, it is a record holder: it is the largest sample ever published of Long Period Variable stars with radial velocity time series.
It should be noted that the knowledge of radial velocities allows us to distinguish between pulsation (so true LPVs) and ellipsoidal variability. 

The tables in the Gaia archive are located in the directory Gaia Focused Product Release and the subdirectory Variability. There are three tables:
\begin{itemize}
    \item \texttt{gaiafpr.vari\_epoch\_radial\_velocity}: Contains the radial velocity time series.
    \item \texttt{gaiafpr.vari\_long\_period\_variable}: Similar to the DR3 table of the same name, but with added radial velocity data (frequencies and amplitudes are recomputed in both the radial velocity and G photometric band time series).
    \item \texttt{gaiafpr.vari\_rad\_vel\_statistics}: Contains statistics on the radial velocities.
\end{itemize}

\subsection{Citizen Science Projet: GaiaVari}
\begin{figure}[htbp]
    \centering
    \includegraphics[width=0.99\linewidth]{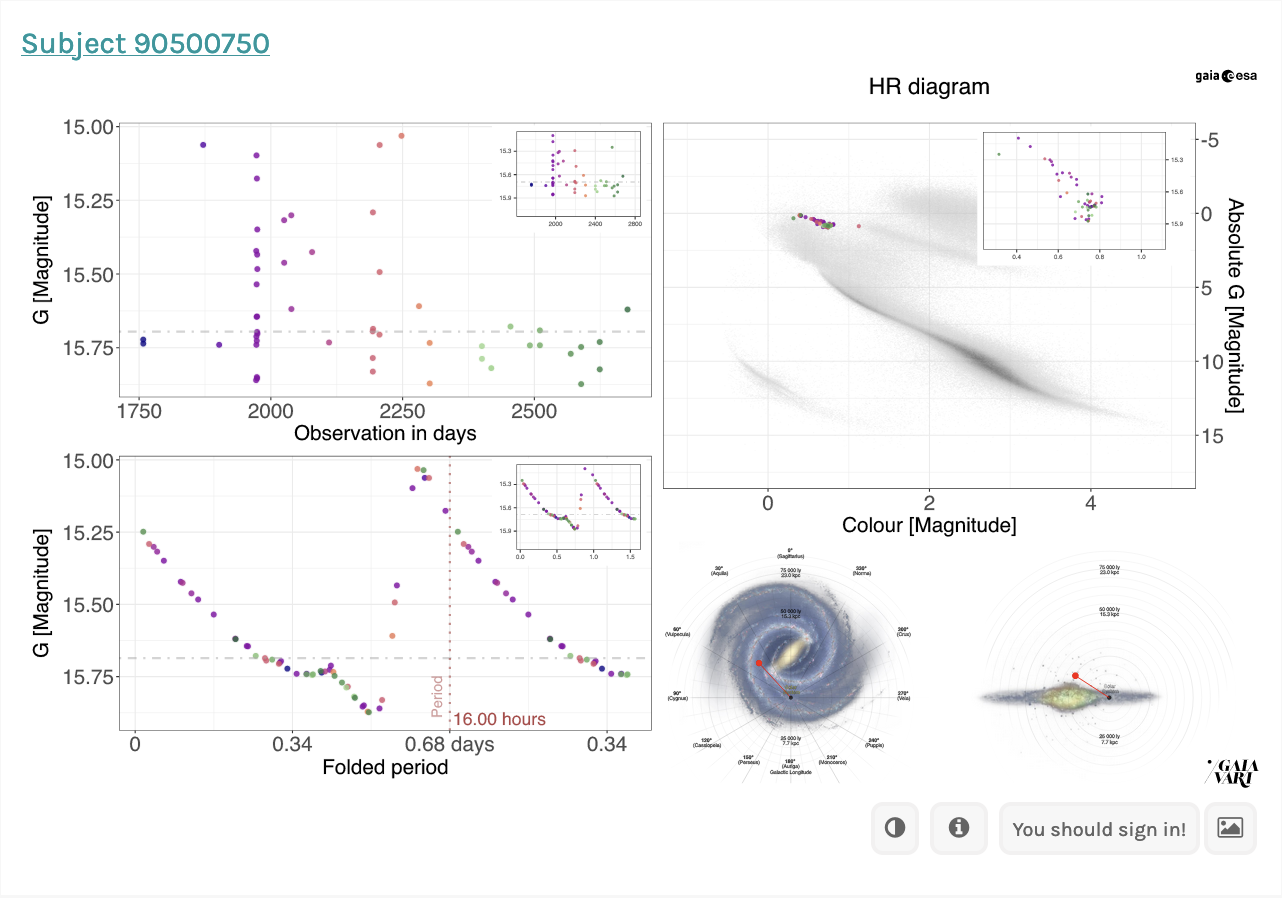} % Replace 'filename' with the actual filename of your image
    \caption{RR Lyrae star of Bailey's type ab in GaiaVari. Upper left panel: the time series: Lower left panel: the folded curve, with the period fo 16 hours. Upper right panel: the motion of the star in the HR diagram. Lower right panel: the position in the Milky Way.}
    \label{fig:GaiaVari}
\end{figure}

%%--------------------
GaiaVari is a citizen science project available on the Zooniverse platform that invites the community to classify variable stars from the public Gaia DR3 results based on the following various visual elements:
\begin{itemize} 
  \item The light curve, representing the $G$ band magnitude measurements as a function of time (colour-coded with time).
  \item The folded curve, displaying the $G$ band magnitude measurements as a function of phase computed from the observed time and an estimated period (the colour-coded according to the light curve, so according to the time, not the phase).
  \item The position in the Hertzsprung-Russell Diagram (the absolute G magnitude as a function of $G_{BP}-G_{RP}$) and the motion of the variable star within this diagram. There are no corrections from the extinction/reddening.
  \item The position in the Milky Way (the distance is taken as the inverse of the parallax).
\end{itemize}
The project is limited to stars with sufficiently good parallax measurements, allowing for the determination of a distance so that it can be placed within the Milky Way.
The sample to classify is not exclusively focused on periodic objects, although the majority of objects chosen fall into this category.
An example of the representation is given in Figure~\ref{fig:GaiaVari} for an RR Lyrae star.

Up to now, there have been one beta campaign and two official campaigns of GaiaVari.
The first one classified the source into the possible following variability types: Eclipsing binaries, Cepheids, RR Lyrae stars and long-period variable stars.
In the second campaign, the following variability types were added to the previous list: $\delta$ Scuti stars and stars with ellipsoidal variability.
In both campaigns, a "None of the above" option was also introduced.

After making a classification choice, and only then, there is the possibility to leave comments.
By choosing this option, one can also discover the Gaia sourceID and the classification from the automated supervised machine learning method \citep{RimoldiniEtal2023}. There is also a link towards ESASky \citep{MerinEtal2015}, and then one can obtain information from databases like Simbad at "Centre de Donn\'ees astronomiques de Strasbourg" (CDS) \citep{Simbad2000}, etc.

Some citizen scientists systematically classify RR Lyrae stars and Cepheids into subgroups, such as RRab, RRc types, or Type-I, Type-II Cepheids.
Such sub-classifications can be compared with the Gaia classification \citep{ClementiniEtal2023, RipepiEtal2023} or the literature.

The community easily identifies the Blazhko effect, although, at times, it is challenging to distinguish Blazhko from an RRd type (double-mode RR Lyrae stars). Additionally, systematic differences in classification between what is published in DR3 and the classifications by some highly educated citizen scientists, particularly in RR Lyrae and Cepheids and their sub-classification, are noteworthy. It will be very interesting to compare these differences.

Due to demand we also created a permanent educational campaign for teachers and schools that can use it as an introduction resource to variable stars and HR diagram understanding.

We plan to conduct additional campaigns. At a point, the plan is to include radial velocity information for a more comprehensive dataset offered by Gaia.

Furthermore, there are plans to write articles on the results of GaiaVari from the first two campaigns, 

GaiaVari is funded by ESA and driven by Sednai Sarl with a collaboration with the University of Geneva, ESA and ScienceNow, see \url{https://www.zooniverse.org/projects/gaia-zooniverse/gaia-vari}.

\section{Conclusions}
%%--------------------
The quantity and quality of Gaia's data are impressive.
Among the many scientific subjects that Gaia is touching, one is particularly positively impacted: time domain astronomy.
 %  ---- Total number of articles ?
In this context, we mostly presented the analysis of photometrically variable sources of DR3, covering 34~months of data with its multitude of results.
Even with the relatively short DR3 time span, numerous records were broken in the number of variable sources, and the unprecedented description of them sets a new standard.
Looking ahead, Gaia's data will nearly double its number of measurements in DR4 and, again, nearly double it for DR5, reaching about 10.5 years of science operational data (if all goes as planned).
The Gaia variability pipeline for processing and analysing data in DR4 and DR5 will receive additional variability classes, features and improvements to the current ones, ensuring not only an increase in quantity but also an enhancement in quality.
This holds the promise of even more interesting results for both the fourth and fifth Gaia data releases.
Undoubtedly, Gaia's data alone provide remarkable insights into variable sources.
However, we can anticipate exciting developments, such as combining Gaia data with other photometric (e.g. Vera Rubin observatory/LSST \citep{IvezicEtal2019}, ZTF \citep{BellmEtal2019}, PLATO \citep{RauerEtal2014}, etc.) or spectroscopic surveys (e.g. 
%4MOST \citep{deJongEtal2019},
WST \citep{MainieriEtal2024}).

% Acknowledgements
% -------------------------
\begin{acknowledgements}
I am grateful to the organisers of this School, in particular to Carine Babusiaux.
I also appreciated the discussions with PhD students and postdocs. I would like to thank Mathieu Van der Swaelmen, whose question on the zero point parallax correction prompted a change in the Gaia consolidated catalogue for DR4. I thank Nicholas Chornay and Krzysztof Nienartowicz for reading and commenting on this manuscript.

I am thankful to the Geneva Gaia team, the Gaia Variability Processing and Analysis international team (CU7), and DPAC consortium. The success of such a mission lies in the competence and dedication of these Consortium members.
Such an article would not exist without their contributions.

During the writing of this document, I had to face the declining health and the passing of my mother, Andr\'ee Eyer P\'etermann. Throughout my life, she was a constant source of support for my high aspirations and my work. I therefore dedicate this text to her.

I met Evry Schatzman only once when I was a young Ph.D. student.
I was deeply moved by his diligence and interest in the young researcher I was, working on Hipparcos variable stars at the time.
This is why I am quite certain that Evry Schatzman would have been most enthusiastic about the Gaia mission and its results.

The texts and figures are original; several software were used to improve the readability and the quality of the text, such as ChatGPT, Grammarly and Google Translate.
This work made use of TOPCAT software \citep{Taylor2005}.
This work has made use of data from the European Space Agency (ESA) mission
{\it Gaia} (\url{https://www.cosmos.esa.int/gaia}), processed by the {\it Gaia}
Data Processing and Analysis Consortium (DPAC,
\url{https://www.cosmos.esa.int/web/gaia/dpac/consortium}). Funding for the DPAC
has been provided by national institutions, in particular, the institutions
participating in the {\it Gaia} Multilateral Agreement.
\end{acknowledgements}

%\appendix
%\section{Appendix}
%The list of the 24 types of variability can be found in Table~\ref{tab:vartypes}. However, some of these are combined types. Therefore, when counting the subclasses, we reach a total of 35 variability types.

%% Bibliography
\bibliographystyle{aa}  % A&A bibliography style file (aa.bst)
\bibliography{eyer_ees2023} % your references in file: Yourfile.bib

\begin{thebibliography}{59}
\expandafter\ifx\csname natexlab\endcsname\relax\def\natexlab#1{#1}\fi

\bibitem[{{Abbott} {et~al.}(2016){Abbott}, {Abbott}, {Abbott}, {Abernathy},
  {Acernese}, {Ackley}, {Adams}, {Adams}, {Addesso}, {Adhikari}, {Adya},
  {Affeldt}, {Agathos}, {Agatsuma}, {Aggarwal}, {Aguiar}, {Aiello}, {Ain},
  {Ajith}, {Allen}, {Allocca}, {Altin}, {Anderson}, {Anderson}, {Arai},
  {Arain}, {Araya}, {Arceneaux}, {Areeda}, {Arnaud}, {Arun}, {Ascenzi},
  {Ashton}, {Ast}, {Aston}, {Astone}, {Aufmuth}, {Aulbert}, {Babak}, {Bacon},
  {Bader}, {Baker}, {Baldaccini}, {Ballardin}, {Ballmer}, {Barayoga},
  {Barclay}, {Barish}, {Barker}, {Barone}, {Barr}, {Barsotti}, {Barsuglia},
  {Barta}, {Bartlett}, {Barton}, {Bartos}, {Bassiri}, {Basti}, {Batch},
  {Baune}, {Bavigadda}, {Bazzan}, {Behnke}, {Bejger}, {Belczynski}, {Bell},
  {Bell}, {Berger}, {Bergman}, {Bergmann}, {Berry}, {Bersanetti}, {Bertolini},
  {Betzwieser}, {Bhagwat}, {Bhandare}, {Bilenko}, {Billingsley}, {Birch},
  {Birney}, {Birnholtz}, {Biscans}, {Bisht}, {Bitossi}, {Biwer}, {Bizouard},
  {Blackburn}, {Blair}, {Blair}, {Blair}, {Bloemen}, {Bock}, {Bodiya}, {Boer},
  {Bogaert}, {Bogan}, {Bohe}, {Bojtos}, {Bond}, {Bondu}, {Bonnand}, {Boom},
  {Bork}, {Boschi}, {Bose}, {Bouffanais}, {Bozzi}, {Bradaschia}, {Brady},
  {Braginsky}, {Branchesi}, {Brau}, {Briant}, {Brillet}, {Brinkmann},
  {Brisson}, {Brockill}, {Brooks}, {Brown}, {Brown}, {Brown}, {Buchanan},
  {Buikema}, {Bulik}, {Bulten}, {Buonanno}, {Buskulic}, {Buy}, {Byer},
  {Cabero}, {Cadonati}, {Cagnoli}, {Cahillane}, {Bustillo}, {Callister},
  {Calloni}, {Camp}, {Cannon}, {Cao}, {Capano}, {Capocasa}, {Carbognani},
  {Caride}, {Casanueva Diaz}, {Casentini}, {Caudill}, {Cavagli{\`a}},
  {Cavalier}, {Cavalieri}, {Cella}, {Cepeda}, {Baiardi}, {Cerretani},
  {Cesarini}, {Chakraborty}, {Chalermsongsak}, {Chamberlin}, {Chan}, {Chao},
  {Charlton}, {Chassande-Mottin}, {Chen}, {Chen}, {Cheng}, {Chincarini},
  {Chiummo}, {Cho}, {Cho}, {Chow}, {Christensen}, {Chu}, {Chua}, {Chung},
  {Ciani}, {Clara}, {Clark}, {Cleva}, {Coccia}, {Cohadon}, {Colla}, {Collette},
  {Cominsky}, {Constancio}, {Conte}, {Conti}, {Cook}, {Corbitt}, {Cornish},
  {Corsi}, {Cortese}, {Costa}, {Coughlin}, {Coughlin}, {Coulon}, {Countryman},
  {Couvares}, {Cowan}, {Coward}, {Cowart}, {Coyne}, {Coyne}, {Craig},
  {Creighton}, {Creighton}, {Cripe}, {Crowder}, {Cruise}, {Cumming},
  {Cunningham}, {Cuoco}, {Dal Canton}, {Danilishin}, {D'Antonio}, {Danzmann},
  {Darman}, {Da Silva Costa}, {Dattilo}, {Dave}, {Daveloza}, {Davier},
  {Davies}, {Daw}, {Day}, {De}, {DeBra}, {Debreczeni}, {Degallaix}, {De
  Laurentis}, {Del{\'e}glise}, {Del Pozzo}, {Denker}, {Dent}, {Dereli},
  {Dergachev}, {DeRosa}, {De Rosa}, {DeSalvo}, {Dhurandhar}, {D{\'\i}az}, {Di
  Fiore}, {Di Giovanni}, {Di Lieto}, {Di Pace}, {Di Palma}, {Di Virgilio},
  {Dojcinoski}, {Dolique}, {Donovan}, {Dooley}, {Doravari}, {Douglas},
  {Downes}, {Drago}, {Drever}, {Driggers}, {Du}, {Ducrot}, {Dwyer}, {Edo},
  {Edwards}, {Effler}, {Eggenstein}, {Ehrens}, {Eichholz}, {Eikenberry},
  {Engels}, {Essick}, {Etzel}, {Evans}, {Evans}, {Everett}, {Factourovich},
  {Fafone}, {Fair}, {Fairhurst}, {Fan}, {Fang}, {Farinon}, {Farr}, {Farr},
  {Favata}, {Fays}, {Fehrmann}, {Fejer}, {Feldbaum}, {Ferrante}, {Ferreira},
  {Ferrini}, {Fidecaro}, {Finn}, {Fiori}, {Fiorucci}, {Fisher}, {Flaminio},
  {Fletcher}, {Fong}, {Fournier}, {Franco}, {Frasca}, {Frasconi}, {Frede},
  {Frei}, {Freise}, {Frey}, {Frey}, {Fricke}, {Fritschel}, {Frolov}, {Fulda},
  {Fyffe}, {Gabbard}, {Gair}, {Gammaitoni}, {Gaonkar}, {Garufi}, {Gatto},
  {Gaur}, {Gehrels}, {Gemme}, {Gendre}, {Genin}, {Gennai}, {George}, {Gergely},
  {Germain}, {Ghosh}, {Ghosh}, {Ghosh}, {Giaime}, {Giardina}, {Giazotto},
  {Gill}, {Glaefke}, {Gleason}, {Goetz}, {Goetz}, {Gondan}, {Gonz{\'a}lez},
  {Castro}, {Gopakumar}, {Gordon}, {Gorodetsky}, {Gossan}, {Gosselin},
  {Gouaty}, {Graef}, {Graff}, {Granata}, {Grant}, {Gras}, {Gray}, {Greco},
  {Green}, {Greenhalgh}, {Groot}, {Grote}, {Grunewald}, {Guidi}, {Guo},
  {Gupta}, {Gupta}, {Gushwa}, {Gustafson}, {Gustafson}, {Hacker}, {Hall},
  {Hall}, {Hammond}, {Haney}, {Hanke}, {Hanks}, {Hanna}, {Hannam}, {Hanson},
  {Hardwick}, {Harms}, {Harry}, {Harry}, {Hart}, {Hartman}, {Haster},
  {Haughian}, {Healy}, {Heefner}, {Heidmann}, {Heintze}, {Heinzel}, {Heitmann},
  {Hello}, {Hemming}, {Hendry}, {Heng}, {Hennig}, {Heptonstall}, {Heurs},
  {Hild}, {Hoak}, {Hodge}, {Hofman}, {Hollitt}, {Holt}, {Holz}, {Hopkins},
  {Hosken}, {Hough}, {Houston}, {Howell}, {Hu}, {Huang}, {Huerta}, {Huet},
  {Hughey}, {Husa}, {Huttner}, {Huynh-Dinh}, {Idrisy}, {Indik}, {Ingram},
  {Inta}, {Isa}, {Isac}, {Isi}, {Islas}, {Isogai}, {Iyer}, {Izumi}, {Jacobson},
  {Jacqmin}, {Jang}, {Jani}, {Jaranowski}, {Jawahar}, {Jim{\'e}nez-Forteza},
  {Johnson}, {Johnson-McDaniel}, {Jones}, {Jones}, {Jonker}, {Ju}, {Haris},
  {Kalaghatgi}, {Kalogera}, {Kandhasamy}, {Kang}, {Kanner}, {Karki},
  {Kasprzack}, {Katsavounidis}, {Katzman}, {Kaufer}, {Kaur}, {Kawabe},
  {Kawazoe}, {K{\'e}f{\'e}lian}, {Kehl}, {Keitel}, {Kelley}, {Kells},
  {Kennedy}, {Keppel}, {Key}, {Khalaidovski}, {Khalili}, {Khan}, {Khan},
  {Khan}, {Khazanov}, {Kijbunchoo}, {Kim}, {Kim}, {Kim}, {Kim}, {Kim}, {Kim},
  {King}, {King}, {Kinzel}, {Kissel}, {Kleybolte}, {Klimenko}, {Koehlenbeck},
  {Kokeyama}, {Koley}, {Kondrashov}, {Kontos}, {Koranda}, {Korobko}, {Korth},
  {Kowalska}, {Kozak}, {Kringel}, {Krishnan}, {Kr{\'o}lak}, {Krueger}, {Kuehn},
  {Kumar}, {Kumar}, {Kuo}, {Kutynia}, {Kwee}, {Lackey}, {Landry}, {Lange},
  {Lantz}, {Lasky}, {Lazzarini}, {Lazzaro}, {Leaci}, {Leavey}, {Lebigot},
  {Lee}, {Lee}, {Lee}, {Lee}, {Lenon}, {Leonardi}, {Leong}, {Leroy},
  {Letendre}, {Levin}, {Levine}, {Li}, {Libson}, {Littenberg}, {Lockerbie},
  {Logue}, {Lombardi}, {London}, {Lord}, {Lorenzini}, {Loriette}, {Lormand},
  {Losurdo}, {Lough}, {Lousto}, {Lovelace}, {L{\"u}ck}, {Lundgren}, {Luo},
  {Lynch}, {Ma}, {MacDonald}, {Machenschalk}, {MacInnis}, {Macleod},
  {Maga{\~n}a-Sandoval}, {Magee}, {Mageswaran}, {Majorana}, {Maksimovic},
  {Malvezzi}, {Man}, {Mandel}, {Mandic}, {Mangano}, {Mansell}, {Manske},
  {Mantovani}, {Marchesoni}, {Marion}, {M{\'a}rka}, {M{\'a}rka}, {Markosyan},
  {Maros}, {Martelli}, {Martellini}, {Martin}, {Martin}, {Martynov}, {Marx},
  {Mason}, {Masserot}, {Massinger}, {Masso-Reid}, {Matichard}, {Matone},
  {Mavalvala}, {Mazumder}, {Mazzolo}, {McCarthy}, {McClelland}, {McCormick},
  {McGuire}, {McIntyre}, {McIver}, {McManus}, {McWilliams}, {Meacher},
  {Meadors}, {Meidam}, {Melatos}, {Mendell}, {Mendoza-Gandara}, {Mercer},
  {Merilh}, {Merzougui}, {Meshkov}, {Messenger}, {Messick}, {Meyers},
  {Mezzani}, {Miao}, {Michel}, {Middleton}, {Mikhailov}, {Milano}, {Miller},
  {Millhouse}, {Minenkov}, {Ming}, {Mirshekari}, {Mishra}, {Mitra},
  {Mitrofanov}, {Mitselmakher}, {Mittleman}, {Moggi}, {Mohan}, {Mohapatra},
  {Montani}, {Moore}, {Moore}, {Moraru}, {Moreno}, {Morriss}, {Mossavi},
  {Mours}, {Mow-Lowry}, {Mueller}, {Mueller}, {Muir}, {Mukherjee}, {Mukherjee},
  {Mukherjee}, {Mukund}, {Mullavey}, {Munch}, {Murphy}, {Murray}, {Mytidis},
  {Nardecchia}, {Naticchioni}, {Nayak}, {Necula}, {Nedkova}, {Nelemans},
  {Neri}, {Neunzert}, {Newton}, {Nguyen}, {Nielsen}, {Nissanke}, {Nitz},
  {Nocera}, {Nolting}, {Normandin}, {Nuttall}, {Oberling}, {Ochsner}, {O'Dell},
  {Oelker}, {Ogin}, {Oh}, {Oh}, {Ohme}, {Oliver}, {Oppermann}, {Oram},
  {O'Reilly}, {O'Shaughnessy}, {Ott}, {Ottaway}, {Ottens}, {Overmier}, {Owen},
  {Pai}, {Pai}, {Palamos}, {Palashov}, {Palomba}, {Pal-Singh}, {Pan}, {Pan},
  {Pankow}, {Pannarale}, {Pant}, {Paoletti}, {Paoli}, {Papa}, {Paris},
  {Parker}, {Pascucci}, {Pasqualetti}, {Passaquieti}, {Passuello},
  {Patricelli}, {Patrick}, {Pearlstone}, {Pedraza}, {Pedurand}, {Pekowsky},
  {Pele}, {Penn}, {Perreca}, {Pfeiffer}, {Phelps}, {Piccinni}, {Pichot},
  {Pickenpack}, {Piergiovanni}, {Pierro}, {Pillant}, {Pinard}, {Pinto},
  {Pitkin}, {Poeld}, {Poggiani}, {Popolizio}, {Post}, {Powell}, {Prasad},
  {Predoi}, {Premachandra}, {Prestegard}, {Price}, {Prijatelj}, {Principe},
  {Privitera}, {Prix}, {Prodi}, {Prokhorov}, {Puncken}, {Punturo}, {Puppo},
  {P{\"u}rrer}, {Qi}, {Qin}, {Quetschke}, {Quintero}, {Quitzow-James}, {Raab},
  {Rabeling}, {Radkins}, {Raffai}, {Raja}, {Rakhmanov}, {Ramet}, {Rapagnani},
  {Raymond}, {Razzano}, {Re}, {Read}, {Reed}, {Regimbau}, {Rei}, {Reid},
  {Reitze}, {Rew}, {Reyes}, {Ricci}, {Riles}, {Robertson}, {Robie}, {Robinet},
  {Rocchi}, {Rolland}, {Rollins}, {Roma}, {Romano}, {Romano}, {Romanov},
  {Romie}, {Rosi{\'n}ska}, {Rowan}, {R{\"u}diger}, {Ruggi}, {Ryan}, {Sachdev},
  {Sadecki}, {Sadeghian}, {Salconi}, {Saleem}, {Salemi}, {Samajdar}, {Sammut},
  {Sampson}, {Sanchez}, {Sandberg}, {Sandeen}, {Sanders}, {Sanders},
  {Sassolas}, {Sathyaprakash}, {Saulson}, {Sauter}, {Savage}, {Sawadsky},
  {Schale}, {Schilling}, {Schmidt}, {Schmidt}, {Schnabel}, {Schofield},
  {Sch{\"o}nbeck}, {Schreiber}, {Schuette}, {Schutz}, {Scott}, {Scott},
  {Sellers}, {Sengupta}, {Sentenac}, {Sequino}, {Sergeev}, {Serna},
  {Setyawati}, {Sevigny}, {Shaddock}, {Shaffer}, {Shah}, {Shahriar}, {Shaltev},
  {Shao}, {Shapiro}, {Shawhan}, {Sheperd}, {Shoemaker}, {Shoemaker}, {Siellez},
  {Siemens}, {Sigg}, {Silva}, {Simakov}, {Singer}, {Singer}, {Singh}, {Singh},
  {Singhal}, {Sintes}, {Slagmolen}, {Smith}, {Smith}, {Smith}, {Smith}, {Son},
  {Sorazu}, {Sorrentino}, {Souradeep}, {Srivastava}, {Staley}, {Steinke},
  {Steinlechner}, {Steinlechner}, {Steinmeyer}, {Stephens}, {Stevenson},
  {Stone}, {Strain}, {Straniero}, {Stratta}, {Strauss}, {Strigin}, {Sturani},
  {Stuver}, {Summerscales}, {Sun}, {Sutton}, {Swinkels}, {Szczepa{\'n}czyk},
  {Tacca}, {Talukder}, {Tanner}, {T{\'a}pai}, {Tarabrin}, {Taracchini},
  {Taylor}, {Theeg}, {Thirugnanasambandam}, {Thomas}, {Thomas}, {Thomas},
  {Thorne}, {Thorne}, {Thrane}, {Tiwari}, {Tiwari}, {Tokmakov}, {Tomlinson},
  {Tonelli}, {Torres}, {Torrie}, {T{\"o}yr{\"a}}, {Travasso}, {Traylor},
  {Trifir{\`o}}, {Tringali}, {Trozzo}, {Tse}, {Turconi}, {Tuyenbayev},
  {Ugolini}, {Unnikrishnan}, {Urban}, {Usman}, {Vahlbruch}, {Vajente},
  {Valdes}, {Vallisneri}, {van Bakel}, {van Beuzekom}, {van den Brand}, {Van
  Den Broeck}, {Vander-Hyde}, {van der Schaaf}, {van Heijningen}, {van Veggel},
  {Vardaro}, {Vass}, {Vas{\'u}th}, {Vaulin}, {Vecchio}, {Vedovato}, {Veitch},
  {Veitch}, {Venkateswara}, {Verkindt}, {Vetrano}, {Vicer{\'e}}, {Vinciguerra},
  {Vine}, {Vinet}, {Vitale}, {Vo}, {Vocca}, {Vorvick}, {Voss}, {Vousden},
  {Vyatchanin}, {Wade}, {Wade}, {Wade}, {Waldman}, {Walker}, {Wallace},
  {Walsh}, {Wang}, {Wang}, {Wang}, {Wang}, {Wang}, {Ward}, {Ward}, {Warner},
  {Was}, {Weaver}, {Wei}, {Weinert}, {Weinstein}, {Weiss}, {Welborn}, {Wen},
  {We{\ss}els}, {Westphal}, {Wette}, {Whelan}, {Whitcomb}, {White}, {Whiting},
  {Wiesner}, {Wilkinson}, {Willems}, {Williams}, {Williams}, {Williamson},
  {Willis}, {Willke}, {Wimmer}, {Winkelmann}, {Winkler}, {Wipf}, {Wiseman},
  {Wittel}, {Woan}, {Worden}, {Wright}, {Wu}, {Yablon}, {Yakushin}, {Yam},
  {Yamamoto}, {Yancey}, {Yap}, {Yu}, {Yvert}, {Zadro{\.Z}ny}, {Zangrando},
  {Zanolin}, {Zendri}, {Zevin}, {Zhang}, {Zhang}, {Zhang}, {Zhang}, {Zhao},
  {Zhou}, {Zhou}, {Zhu}, {Zucker}, {Zuraw}, {Zweizig}, {LIGO Scientific
  Collaboration}, \& {Virgo Collaboration}}]{2016PhRvL.116f1102A}
{Abbott}, B.~P., {Abbott}, R., {Abbott}, T.~D., {et~al.} 2016, \prl, 116,
  061102

\bibitem[{{Appenzeller}(2013)}]{Apprenzeller2013}
{Appenzeller}, I. 2013, {Introduction to Astronomical Spectroscopy}

\bibitem[{{Bakos} {et~al.}(2004){Bakos}, {Noyes}, {Kov{\'a}cs}, {Stanek},
  {Sasselov}, \& {Domsa}}]{HAT2004}
{Bakos}, G., {Noyes}, R.~W., {Kov{\'a}cs}, G., {et~al.} 2004, \pasp, 116, 266

\bibitem[{{Bellm} {et~al.}(2019){Bellm}, {Kulkarni}, {Graham}, {Dekany},
  {Smith}, {Riddle}, {Masci}, {Helou}, {Prince}, {Adams}, {Barbarino},
  {Barlow}, {Bauer}, {Beck}, {Belicki}, {Biswas}, {Blagorodnova}, {Bodewits},
  {Bolin}, {Brinnel}, {Brooke}, {Bue}, {Bulla}, {Burruss}, {Cenko}, {Chang},
  {Connolly}, {Coughlin}, {Cromer}, {Cunningham}, {De}, {Delacroix}, {Desai},
  {Duev}, {Eadie}, {Farnham}, {Feeney}, {Feindt}, {Flynn}, {Franckowiak},
  {Frederick}, {Fremling}, {Gal-Yam}, {Gezari}, {Giomi}, {Goldstein},
  {Golkhou}, {Goobar}, {Groom}, {Hacopians}, {Hale}, {Henning}, {Ho}, {Hover},
  {Howell}, {Hung}, {Huppenkothen}, {Imel}, {Ip}, {Ivezi{\'c}}, {Jackson},
  {Jones}, {Juric}, {Kasliwal}, {Kaspi}, {Kaye}, {Kelley}, {Kowalski},
  {Kramer}, {Kupfer}, {Landry}, {Laher}, {Lee}, {Lin}, {Lin}, {Lunnan},
  {Giomi}, {Mahabal}, {Mao}, {Miller}, {Monkewitz}, {Murphy}, {Ngeow},
  {Nordin}, {Nugent}, {Ofek}, {Patterson}, {Penprase}, {Porter}, {Rauch},
  {Rebbapragada}, {Reiley}, {Rigault}, {Rodriguez}, {van Roestel}, {Rusholme},
  {van Santen}, {Schulze}, {Shupe}, {Singer}, {Soumagnac}, {Stein}, {Surace},
  {Sollerman}, {Szkody}, {Taddia}, {Terek}, {Van Sistine}, {van Velzen},
  {Vestrand}, {Walters}, {Ward}, {Ye}, {Yu}, {Yan}, \&
  {Zolkower}}]{BellmEtal2019}
{Bellm}, E.~C., {Kulkarni}, S.~R., {Graham}, M.~J., {et~al.} 2019, \pasp, 131,
  018002

\bibitem[{{Borucki} {et~al.}(2010){Borucki}, {Koch}, {Basri}, {Batalha},
  {Brown}, {Caldwell}, {Caldwell}, {Christensen-Dalsgaard}, {Cochran},
  {DeVore}, {Dunham}, {Dupree}, {Gautier}, {Geary}, {Gilliland}, {Gould},
  {Howell}, {Jenkins}, {Kondo}, {Latham}, {Marcy}, {Meibom}, {Kjeldsen},
  {Lissauer}, {Monet}, {Morrison}, {Sasselov}, {Tarter}, {Boss}, {Brownlee},
  {Owen}, {Buzasi}, {Charbonneau}, {Doyle}, {Fortney}, {Ford}, {Holman},
  {Seager}, {Steffen}, {Welsh}, {Rowe}, {Anderson}, {Buchhave}, {Ciardi},
  {Walkowicz}, {Sherry}, {Horch}, {Isaacson}, {Everett}, {Fischer}, {Torres},
  {Johnson}, {Endl}, {MacQueen}, {Bryson}, {Dotson}, {Haas}, {Kolodziejczak},
  {Van Cleve}, {Chandrasekaran}, {Twicken}, {Quintana}, {Clarke}, {Allen},
  {Li}, {Wu}, {Tenenbaum}, {Verner}, {Bruhweiler}, {Barnes}, \&
  {Prsa}}]{BoruckiEtal2010}
{Borucki}, W.~J., {Koch}, D., {Basri}, G., {et~al.} 2010, Science, 327, 977

\bibitem[{{Burki} {et~al.}(1995){Burki}, {Rufener}, {Burnet}, {Richard},
  {Blecha}, \& {Bratschi}}]{BurkiEtal1995}
{Burki}, G., {Rufener}, F., {Burnet}, M., {et~al.} 1995, \aaps, 112, 383

\bibitem[{{Carnerero} {et~al.}(2023){Carnerero}, {Raiteri}, {Rimoldini},
  {Busonero}, {Licata}, {Mowlavi}, {Lecoeur-Ta{\"\i}bi}, {Audard}, {Holl},
  {Gavras}, {Nienartowicz}, {Jevardat de Fombelle}, {Carballo}, {Clementini},
  {Delchambre}, {Klioner}, {Lattanzi}, \& {Eyer}}]{CarnereroEtal2023}
{Carnerero}, M.~I., {Raiteri}, C.~M., {Rimoldini}, L., {et~al.} 2023, \aap,
  674, A24

\bibitem[{{Carney} {et~al.}(1992){Carney}, {Storm}, \&
  {Jones}}]{CarneyEtal1992}
{Carney}, B.~W., {Storm}, J., \& {Jones}, R.~V. 1992, \apj, 386, 663

\bibitem[{{Chakrabarti} {et~al.}(2023){Chakrabarti}, {Simon}, {Craig},
  {Reggiani}, {Brandt}, {Guhathakurta}, {Dalba}, {Kirby}, {Chang}, {Hey},
  {Savino}, {Geha}, \& {Thompson}}]{ChakrabartiEtal2023}
{Chakrabarti}, S., {Simon}, J.~D., {Craig}, P.~A., {et~al.} 2023, \aj, 166, 6

\bibitem[{{Clementini} {et~al.}(2023){Clementini}, {Ripepi}, {Garofalo},
  {Molinaro}, {Muraveva}, {Leccia}, {Rimoldini}, {Holl}, {Jevardat de
  Fombelle}, {Sartoretti}, {Marchal}, {Audard}, {Nienartowicz}, {Andrae},
  {Marconi}, {Szabados}, {Evans}, {Lecoeur-Taibi}, {Mowlavi}, {Musella}, \&
  {Eyer}}]{ClementiniEtal2023}
{Clementini}, G., {Ripepi}, V., {Garofalo}, A., {et~al.} 2023, \aap, 674, A18

\bibitem[{{Clementini} {et~al.}(2016){Clementini}, {Ripepi}, {Leccia},
  {Mowlavi}, {Lecoeur-Taibi}, {Marconi}, {Szabados}, {Eyer}, {Guy},
  {Rimoldini}, {Jevardat de Fombelle}, {Holl}, {Busso}, {Charnas}, {Cuypers},
  {De Angeli}, {De Ridder}, {Debosscher}, {Evans}, {Klagyivik}, {Musella},
  {Nienartowicz}, {Ord{\'o}{\~n}ez}, {Regibo}, {Riello}, {Sarro}, \&
  {S{\"u}veges}}]{ClementiniEtal2016}
{Clementini}, G., {Ripepi}, V., {Leccia}, S., {et~al.} 2016, \aap, 595, A133

\bibitem[{Davis {et~al.}(1968)Davis, Harmer, \& Hoffman}]{Davies1968}
Davis, R., Harmer, D.~S., \& Hoffman, K.~C. 1968, Phys. Rev. Lett., 20, 1205

\bibitem[{{de Jong} {et~al.}(2019){de Jong}, {Agertz}, {Berbel}, {Aird},
  {Alexander}, {Amarsi}, {Anders}, {Andrae}, {Ansarinejad}, {Ansorge},
  {Antilogus}, {Anwand-Heerwart}, {Arentsen}, {Arnadottir}, {Asplund}, {Auger},
  {Azais}, {Baade}, {Baker}, {Baker}, {Balbinot}, {Baldry}, {Banerji},
  {Barden}, {Barklem}, {Barth{\'e}l{\'e}my-Mazot}, {Battistini}, {Bauer},
  {Bell}, {Bellido-Tirado}, {Bellstedt}, {Belokurov}, {Bensby}, {Bergemann},
  {Bestenlehner}, {Bielby}, {Bilicki}, {Blake}, {Bland-Hawthorn}, {Boeche},
  {Boland}, {Boller}, {Bongard}, {Bongiorno}, {Bonifacio}, {Boudon}, {Brooks},
  {Brown}, {Brown}, {Br{\"u}ggen}, {Brynnel}, {Brzeski}, {Buchert},
  {Buschkamp}, {Caffau}, {Caillier}, {Carrick}, {Casagrande}, {Case}, {Casey},
  {Cesarini}, {Cescutti}, {Chapuis}, {Chiappini}, {Childress}, {Christlieb},
  {Church}, {Cioni}, {Cluver}, {Colless}, {Collett}, {Comparat}, {Cooper},
  {Couch}, {Courbin}, {Croom}, {Croton}, {Daguis{\'e}}, {Dalton}, {Davies},
  {Davis}, {de Laverny}, {Deason}, {Dionies}, {Disseau}, {Doel}, {D{\"o}scher},
  {Driver}, {Dwelly}, {Eckert}, {Edge}, {Edvardsson}, {Youssoufi}, {Elhaddad},
  {Enke}, {Erfanianfar}, {Farrell}, {Fechner}, {Feiz}, {Feltzing}, {Ferreras},
  {Feuerstein}, {Feuillet}, {Finoguenov}, {Ford}, {Fotopoulou}, {Fouesneau},
  {Frenk}, {Frey}, {Gaessler}, {Geier}, {Gentile Fusillo}, {Gerhard},
  {Giannantonio}, {Giannone}, {Gibson}, {Gillingham},
  {Gonz{\'a}lez-Fern{\'a}ndez}, {Gonzalez-Solares}, {Gottloeber}, {Gould},
  {Grebel}, {Gueguen}, {Guiglion}, {Haehnelt}, {Hahn}, {Hansen}, {Hartman},
  {Hauptner}, {Hawkins}, {Haynes}, {Haynes}, {Heiter}, {Helmi}, {Aguayo},
  {Hewett}, {Hinton}, {Hobbs}, {Hoenig}, {Hofman}, {Hook}, {Hopgood},
  {Hopkins}, {Hourihane}, {Howes}, {Howlett}, {Huet}, {Irwin}, {Iwert},
  {Jablonka}, {Jahn}, {Jahnke}, {Jarno}, {Jin}, {Jofre}, {Johl}, {Jones},
  {J{\"o}nsson}, {Jordan}, {Karovicova}, {Khalatyan}, {Kelz}, {Kennicutt},
  {King}, {Kitaura}, {Klar}, {Klauser}, {Kneib}, {Koch}, {Koposov},
  {Kordopatis}, {Korn}, {Kosmalski}, {Kotak}, {Kovalev}, {Kreckel}, {Kripak},
  {Krumpe}, {Kuijken}, {Kunder}, {Kushniruk}, {Lam}, {Lamer}, {Laurent},
  {Lawrence}, {Lehmitz}, {Lemasle}, {Lewis}, {Li}, {Lidman}, {Lind}, {Liske},
  {Lizon}, {Loveday}, {Ludwig}, {McDermid}, {Maguire}, {Mainieri}, {Mali},
  {Mandel}, {Mandel}, {Mannering}, {Martell}, {Martinez Delgado}, {Matijevic},
  {McGregor}, {McMahon}, {McMillan}, {Mena}, {Merloni}, {Meyer}, {Michel},
  {Micheva}, {Migniau}, {Minchev}, {Monari}, {Muller}, {Murphy},
  {Muthukrishna}, {Nandra}, {Navarro}, {Ness}, {Nichani}, {Nichol}, {Nicklas},
  {Niederhofer}, {Norberg}, {Obreschkow}, {Oliver}, {Owers}, {Pai},
  {Pankratow}, {Parkinson}, {Paschke}, {Paterson}, {Pecontal}, {Parry},
  {Phillips}, {Pillepich}, {Pinard}, {Pirard}, {Piskunov}, {Plank},
  {Pl{\"u}schke}, {Pons}, {Popesso}, {Power}, {Pragt}, {Pramskiy}, {Pryer},
  {Quattri}, {Queiroz}, {Quirrenbach}, {Rahurkar}, {Raichoor}, {Ramstedt},
  {Rau}, {Recio-Blanco}, {Reiss}, {Renaud}, {Revaz}, {Rhode}, {Richard},
  {Richter}, {Rix}, {Robotham}, {Roelfsema}, {Romaniello}, {Rosario},
  {Rothmaier}, {Roukema}, {Ruchti}, {Rupprecht}, {Rybizki}, {Ryde}, {Saar},
  {Sadler}, {Sahl{\'e}n}, {Salvato}, {Sassolas}, {Saunders}, {Saviauk},
  {Sbordone}, {Schmidt}, {Schnurr}, {Scholz}, {Schwope}, {Seifert}, {Shanks},
  {Sheinis}, {Sivov}, {Sk{\'u}lad{\'o}ttir}, {Smartt}, {Smedley}, {Smith},
  {Smith}, {Sorce}, {Spitler}, {Starkenburg}, {Steinmetz}, {Stilz}, {Storm},
  {Sullivan}, {Sutherland}, {Swann}, {Tamone}, {Taylor}, {Teillon}, {Tempel},
  {ter Horst}, {Thi}, {Tolstoy}, {Trager}, {Traven}, {Tremblay}, {Tresse},
  {Valentini}, {van de Weygaert}, {van den Ancker}, {Veljanoski}, {Venkatesan},
  {Wagner}, {Wagner}, {Walcher}, {Waller}, {Walton}, {Wang}, {Winkler},
  {Wisotzki}, {Worley}, {Worseck}, {Xiang}, {Xu}, {Yong}, {Zhao}, {Zheng},
  {Zscheyge}, \& {Zucker}}]{deJongEtal2019}
{de Jong}, R.~S., {Agertz}, O., {Berbel}, A.~A., {et~al.} 2019, The Messenger,
  175, 3

\bibitem[{{Distefano} {et~al.}(2023){Distefano}, {Lanzafame}, {Brugaletta},
  {Holl}, {Lanza}, {Messina}, {Pagano}, {Audard}, {Jevardat de Fombelle},
  {Lecoeur-Taibi}, {Mowlavi}, {Nienartowicz}, {Rimoldini}, {Evans}, {Riello},
  {Garc{\'\i}a-Lario}, {Gavras}, \& {Eyer}}]{DistefanoEtal2023}
{Distefano}, E., {Lanzafame}, A.~C., {Brugaletta}, E., {et~al.} 2023, \aap,
  674, A20

\bibitem[{{El-Badry} {et~al.}(2023){El-Badry}, {Rix}, {Quataert}, {Howard},
  {Isaacson}, {Fuller}, {Hawkins}, {Breivik}, {Wong}, {Rodriguez}, {Conroy},
  {Shahaf}, {Mazeh}, {Arenou}, {Burdge}, {Bashi}, {Faigler}, {Weisz},
  {Seeburger}, {Almada Monter}, \& {Wojno}}]{El-BadryEtal2023}
{El-Badry}, K., {Rix}, H.-W., {Quataert}, E., {et~al.} 2023, \mnras, 518, 1057

\bibitem[{{Evans} {et~al.}(2023){Evans}, {Eyer}, {Busso}, {Riello}, {De
  Angeli}, {Burgess}, {Audard}, {Clementini}, {Garofalo}, {Holl}, {Jevardat de
  Fombelle}, {Lanzafame}, {Lecoeur-Taibi}, {Mowlavi}, {Nienartowicz},
  {Palaversa}, \& {Rimoldini}}]{EvansEtal2023}
{Evans}, D.~W., {Eyer}, L., {Busso}, G., {et~al.} 2023, \aap, 674, A4

\bibitem[{{Eyer}(2006)}]{Eyer2006}
{Eyer}, L. 2006, \memsai, 77, 549

\bibitem[{{Eyer} {et~al.}(2023){Eyer}, {Audard}, {Holl}, {Rimoldini},
  {Carnerero}, {Clementini}, {De Ridder}, {Distefano}, {Evans}, {Gavras},
  {Gomel}, {Lebzelter}, {Marton}, {Mowlavi}, {Panahi}, {Ripepi}, {Wyrzykowski},
  {Nienartowicz}, {Jevardat de Fombelle}, {Lecoeur-Taibi}, {Rohrbasser},
  {Riello}, {Garc{\'\i}a-Lario}, {Lanzafame}, {Mazeh}, {Raiteri}, {Zucker},
  {{\'A}brah{\'a}m}, {Aerts}, {Aguado}, {Anderson}, {Bashi}, {Binnenfeld},
  {Faigler}, {Garofalo}, {Karbevska}, {K{\'o}sp{\'a}l}, {Kruszy{\'n}ska},
  {Kun}, {Lanza}, {Leccia}, {Marconi}, {Messina}, {Molinaro}, {Moln{\'a}r},
  {Muraveva}, {Musella}, {Nagy}, {Pagano}, {Palaversa}, {Plachy}, {Pr{\v{s}}a},
  {Rybicki}, {Shahaf}, {Szabados}, {Szegedi-Elek}, {Trabucchi}, {Barblan},
  {Grenon}, {Roelens}, \& {S{\"u}veges}}]{EyerEtal2023}
{Eyer}, L., {Audard}, M., {Holl}, B., {et~al.} 2023, \aap, 674, A13

\bibitem[{{Eyer} \& {Mignard}(2005)}]{EyerMignard2005}
{Eyer}, L. \& {Mignard}, F. 2005, \mnras, 361, 1136

\bibitem[{{Eyer} {et~al.}(2017){Eyer}, {Mowlavi}, {Evans}, {Nienartowicz},
  {Ordonez}, {Holl}, {Lecoeur-Taibi}, {Riello}, {Clementini}, {Cuypers}, {De
  Ridder}, {Lanzafame}, {Sarro}, {Charnas}, {Guy}, {Jevardat de Fombelle},
  {Rimoldini}, {S{\"u}veges}, {Mignard}, {Busso}, {De Angeli}, {van Leeuwen},
  {Dubath}, {Beck}, {Aguado}, {Debosscher}, {Distefano}, {Fuchs}, {Koubsky},
  {Lebzelter}, {Leccia}, {Lopez}, {Moitinho}, {Regibo}, {Ripepi}, {Roelens},
  {Szabados}, {Tingley}, {Votruba}, {Zucker}, {Aerts}, {Barblan},
  {Blanco-Cuaresma}, {Grenon}, {Jan}, {Lorenz}, {Miranda}, {Morgenthaler},
  {Ordenovic}, {Palaversa}, {Prsa}, {Ruiz-Fuertes}, {Anderson}, {Delgado},
  {Dzigan}, {Hudec}, {Jonckheere}, {Klagyivik}, {Kutka}, {Moniez}, {Nicoletti},
  {Park}, {Van Hemelryck}, {Varadi}, {Kochoska}, {Lanza}, {Marconi},
  {Marschalko}, {Messina}, {Musella}, {Pagano}, {Sadowski}, \&
  {Schultheis}}]{EyerEtal2017}
{Eyer}, L., {Mowlavi}, N., {Evans}, D.~W., {et~al.} 2017, arXiv e-prints,
  arXiv:1702.03295

\bibitem[{{Eyer} {et~al.}(2015){Eyer}, {Rimoldini}, {Holl}, {North}, {Zucker},
  {Evans}, {Pourbaix}, {Hodgkin}, {Thuillot}, {Mowlavi}, \&
  {Carry}}]{EyerEtal2015}
{Eyer}, L., {Rimoldini}, L., {Holl}, B., {et~al.} 2015, in Astronomical Society
  of the Pacific Conference Series, Vol. 496, Living Together: Planets, Host
  Stars and Binaries, ed. S.~M. {Rucinski}, G.~{Torres}, \& M.~{Zejda}, 121

\bibitem[{{Gaia Collaboration} {et~al.}(2017){Gaia Collaboration},
  {Clementini}, {Eyer}, {Ripepi}, {Marconi}, {Muraveva}, {Garofalo}, {Sarro},
  {Palmer}, {Luri}, {Molinaro}, {Rimoldini}, {Szabados}, {Musella}, {Anderson},
  {Prusti}, {de Bruijne}, {Brown}, {Vallenari}, {Babusiaux}, {Bailer-Jones},
  {Bastian}, {Biermann}, {Evans}, {Jansen}, {Jordi}, {Klioner}, {Lammers},
  {Lindegren}, {Mignard}, {Panem}, {Pourbaix}, {Randich}, {Sartoretti},
  {Siddiqui}, {Soubiran}, {Valette}, {van Leeuwen}, {Walton}, {Aerts},
  {Arenou}, {Cropper}, {Drimmel}, {H{\o}g}, {Katz}, {Lattanzi}, {O'Mullane},
  {Grebel}, {Holland}, {Huc}, {Passot}, {Perryman}, {Bramante}, {Cacciari},
  {Casta{\~n}eda}, {Chaoul}, {Cheek}, {De Angeli}, {Fabricius}, {Guerra},
  {Hern{\'a}ndez}, {Jean-Antoine-Piccolo}, {Masana}, {Messineo}, {Mowlavi},
  {Nienartowicz}, {Ord{\'o}{\~n}ez-Blanco}, {Panuzzo}, {Portell}, {Richards},
  {Riello}, {Seabroke}, {Tanga}, {Th{\'e}venin}, {Torra}, {Els},
  {Gracia-Abril}, {Comoretto}, {Garcia-Reinaldos}, {Lock}, {Mercier},
  {Altmann}, {Andrae}, {Astraatmadja}, {Bellas-Velidis}, {Benson}, {Berthier},
  {Blomme}, {Busso}, {Carry}, {Cellino}, {Cowell}, {Creevey}, {Cuypers},
  {Davidson}, {De Ridder}, {de Torres}, {Delchambre}, {Dell'Oro}, {Ducourant},
  {Fr{\'e}mat}, {Garc{\'\i}a-Torres}, {Gosset}, {Halbwachs}, {Hambly},
  {Harrison}, {Hauser}, {Hestroffer}, {Hodgkin}, {Huckle}, {Hutton},
  {Jasniewicz}, {Jordan}, {Kontizas}, {Korn}, {Lanzafame}, {Manteiga},
  {Moitinho}, {Muinonen}, {Osinde}, {Pancino}, {Pauwels}, {Petit},
  {Recio-Blanco}, {Robin}, {Siopis}, {Smith}, {Smith}, {Sozzetti}, {Thuillot},
  {van Reeven}, {Viala}, {Abbas}, {Abreu Aramburu}, {Accart}, {Aguado},
  {Allan}, {Allasia}, {Altavilla}, {{\'A}lvarez}, {Alves}, {Andrei}, {Anglada
  Varela}, {Antiche}, {Antoja}, {Ant{\'o}n}, {Arcay}, {Bach}, {Baker},
  {Balaguer-N{\'u}{\~n}ez}, {Barache}, {Barata}, {Barbier}, {Barblan}, {Barrado
  y Navascu{\'e}s}, {Barros}, {Barstow}, {Becciani}, {Bellazzini}, {Bello
  Garc{\'\i}a}, {Belokurov}, {Bendjoya}, {Berihuete}, {Bianchi},
  {Bienaym{\'e}}, {Billebaud}, {Blagorodnova}, {Blanco-Cuaresma}, {Boch},
  {Bombrun}, {Borrachero}, {Bouquillon}, {Bourda}, {Bragaglia}, {Breddels},
  {Brouillet}, {Br{\"u}semeister}, {Bucciarelli}, {Burgess}, {Burgon},
  {Burlacu}, {Busonero}, {Buzzi}, {Caffau}, {Cambras}, {Campbell},
  {Cancelliere}, {Cantat-Gaudin}, {Carlucci}, {Carrasco}, {Castellani},
  {Charlot}, {Charnas}, {Chiavassa}, {Clotet}, {Cocozza}, {Collins},
  {Costigan}, {Crifo}, {Cross}, {Crosta}, {Crowley}, {Dafonte}, {Damerdji},
  {Dapergolas}, {David}, {David}, {De Cat}, {de Felice}, {de Laverny}, {De
  Luise}, {De March}, {de Souza}, {Debosscher}, {del Pozo}, {Delbo}, {Delgado},
  {Delgado}, {Di Matteo}, {Diakite}, {Distefano}, {Dolding}, {Dos Anjos},
  {Drazinos}, {Dur{\'a}n}, {Dzigan}, {Edvardsson}, {Enke}, {Evans}, {Eynard
  Bontemps}, {Fabre}, {Fabrizio}, {Falc{\~a}o}, {Farr{\`a}s Casas}, {Federici},
  {Fedorets}, {Fern{\'a}ndez-Hern{\'a}ndez}, {Fernique}, {Fienga}, {Figueras},
  {Filippi}, {Findeisen}, {Fonti}, {Fouesneau}, {Fraile}, {Fraser}, {Fuchs},
  {Gai}, {Galleti}, {Galluccio}, {Garabato}, {Garc{\'\i}a-Sedano}, {Garralda},
  {Gavras}, {Gerssen}, {Geyer}, {Gilmore}, {Girona}, {Giuffrida}, {Gomes},
  {Gonz{\'a}lez-Marcos}, {Gonz{\'a}lez-N{\'u}{\~n}ez}, {Gonz{\'a}lez-Vidal},
  {Granvik}, {Guerrier}, {Guillout}, {Guiraud}, {G{\'u}rpide},
  {Guti{\'e}rrez-S{\'a}nchez}, {Guy}, {Haigron}, {Hatzidimitriou}, {Haywood},
  {Heiter}, {Helmi}, {Hobbs}, {Hofmann}, {Holl}, {Holland}, {Hunt}, {Hypki},
  {Icardi}, {Irwin}, {Jevardat de Fombelle}, {Jofr{\'e}}, {Jonker}, {Jorissen},
  {Julbe}, {Karampelas}, {Kochoska}, {Kohley}, {Kolenberg}, {Kontizas},
  {Koposov}, {Kordopatis}, {Koubsky}, {Krone-Martins}, {Kudryashova},
  {Bachchan}, {Lacoste-Seris}, {Lanza}, {Lavigne}, {Le Poncin-Lafitte},
  {Lebreton}, {Lebzelter}, {Leccia}, {Leclerc}, {Lecoeur-Taibi}, {Lemaitre},
  {Lenhardt}, {Leroux}, {Liao}, {Licata}, {Lindstr{\o}m}, {Lister}, {Livanou},
  {Lobel}, {L{\"o}ffler}, {L{\'o}pez}, {Lorenz}, {MacDonald}, {Magalh{\~a}es
  Fernandes}, {Managau}, {Mann}, {Mantelet}, {Marchal}, {Marchant}, {Marinoni},
  {Marrese}, {Marschalk{\'o}}, {Marshall}, {Mart{\'\i}n-Fleitas}, {Martino},
  {Mary}, {Matijevi{\v{c}}}, {McMillan}, {Messina}, {Michalik}, {Millar},
  {Miranda}, {Molina}, {Molinaro}, {Moln{\'a}r}, {Moniez}, {Montegriffo},
  {Mor}, {Mora}, {Morbidelli}, {Morel}, {Morgenthaler}, {Morris}, {Mulone},
  {Narbonne}, {Nelemans}, {Nicastro}, {Noval}, {Ord{\'e}novic},
  {Ordieres-Mer{\'e}}, {Osborne}, {Pagani}, {Pagano}, {Pailler}, {Palacin},
  {Palaversa}, {Parsons}, {Pecoraro}, {Pedrosa}, {Pentik{\"a}inen}, {Pichon},
  {Piersimoni}, {Pineau}, {Plachy}, {Plum}, {Poujoulet}, {Pr{\v{s}}a},
  {Pulone}, {Ragaini}, {Rago}, {Rambaux}, {Ramos-Lerate}, {Ranalli}, {Rauw},
  {Read}, {Regibo}, {Reyl{\'e}}, {Ribeiro}, {Riva}, {Rixon}, {Roelens},
  {Romero-G{\'o}mez}, {Rowell}, {Royer}, {Ruiz-Dern}, {Sadowski}, {Sagrist{\`a}
  Sell{\'e}s}, {Sahlmann}, {Salgado}, {Salguero}, {Sarasso}, {Savietto},
  {Schultheis}, {Sciacca}, {Segol}, {Segovia}, {Segransan}, {Shih},
  {Smareglia}, {Smart}, {Solano}, {Solitro}, {Sordo}, {Soria Nieto}, {Souchay},
  {Spagna}, {Spoto}, {Stampa}, {Steele}, {Steidelm{\"u}ller}, {Stephenson},
  {Stoev}, {Suess}, {S{\"u}veges}, {Surdej}, {Szegedi-Elek}, {Tapiador},
  {Taris}, {Tauran}, {Taylor}, {Teixeira}, {Terrett}, {Tingley}, {Trager},
  {Turon}, {Ulla}, {Utrilla}, {Valentini}, {van Elteren}, {Van Hemelryck}, {van
  Leeuwen}, {Varadi}, {Vecchiato}, {Veljanoski}, {Via}, {Vicente}, {Vogt},
  {Voss}, {Votruba}, {Voutsinas}, {Walmsley}, {Weiler}, {Weingrill}, {Wevers},
  {Wyrzykowski}, {Yoldas}, {{\v{Z}}erjal}, {Zucker}, {Zurbach}, {Zwitter},
  {Alecu}, {Allen}, {Allende Prieto}, {Amorim}, {Anglada-Escud{\'e}},
  {Arsenijevic}, {Azaz}, {Balm}, {Beck}, {Bernstein}, {Bigot}, {Bijaoui},
  {Blasco}, {Bonfigli}, {Bono}, {Boudreault}, {Bressan}, {Brown}, {Brunet},
  {Bunclark}, {Buonanno}, {Butkevich}, {Carret}, {Carrion}, {Chemin},
  {Ch{\'e}reau}, {Corcione}, {Darmigny}, {de Boer}, {de Teodoro}, {de Zeeuw},
  {Delle Luche}, {Domingues}, {Dubath}, {Fodor}, {Fr{\'e}zouls}, {Fries},
  {Fustes}, {Fyfe}, {Gallardo}, {Gallegos}, {Gardiol}, {Gebran}, {Gomboc},
  {G{\'o}mez}, {Grux}, {Gueguen}, {Heyrovsky}, {Hoar}, {Iannicola}, {Isasi
  Parache}, {Janotto}, {Joliet}, {Jonckheere}, {Keil}, {Kim}, {Klagyivik},
  {Klar}, {Knude}, {Kochukhov}, {Kolka}, {Kos}, {Kutka}, {Lainey}, {LeBouquin},
  {Liu}, {Loreggia}, {Makarov}, {Marseille}, {Martayan}, {Martinez-Rubi},
  {Massart}, {Meynadier}, {Mignot}, {Munari}, {Nguyen}, {Nordlander},
  {O'Flaherty}, {Ocvirk}, {Olias Sanz}, {Ortiz}, {Osorio}, {Oszkiewicz},
  {Ouzounis}, {Park}, {Pasquato}, {Peltzer}, {Peralta}, {P{\'e}turaud},
  {Pieniluoma}, {Pigozzi}, {Poels}, {Prat}, {Prod'homme}, {Raison}, {Rebordao},
  {Risquez}, {Rocca-Volmerange}, {Rosen}, {Ruiz-Fuertes}, {Russo}, {Serraller
  Vizcaino}, {Short}, {Siebert}, {Silva}, {Sinachopoulos}, {Slezak}, {Soffel},
  {Sosnowska}, {Strai{\v{z}}ys}, {ter Linden}, {Terrell}, {Theil}, {Tiede},
  {Troisi}, {Tsalmantza}, {Tur}, {Vaccari}, {Vachier}, {Valles}, {Van Hamme},
  {Veltz}, {Virtanen}, {Wallut}, {Wichmann}, {Wilkinson}, {Ziaeepour}, \&
  {Zschocke}}]{ClementiniEtal2017}
{Gaia Collaboration}, {Clementini}, G., {Eyer}, L., {et~al.} 2017, \aap, 605,
  A79

\bibitem[{{Gaia Collaboration} {et~al.}(2023{\natexlab{a}}){Gaia
  Collaboration}, {De Ridder}, {Ripepi}, {Aerts}, {Palaversa}, {Eyer}, {Holl},
  {Audard}, {Rimoldini}, {Brown}, {Vallenari}, {Prusti}, {de Bruijne},
  {Arenou}, {Babusiaux}, {Biermann}, {Creevey}, {Ducourant}, {Evans}, {Guerra},
  {Hutton}, {Jordi}, {Klioner}, {Lammers}, {Lindegren}, {Luri}, {Mignard},
  {Panem}, {Pourbaix}, {Randich}, {Sartoretti}, {Soubiran}, {Tanga}, {Walton},
  {Bailer-Jones}, {Bastian}, {Drimmel}, {Jansen}, {Katz}, {Lattanzi}, {van
  Leeuwen}, {Bakker}, {Cacciari}, {Casta{\~n}eda}, {De Angeli}, {Fabricius},
  {Fouesneau}, {Fr{\'e}mat}, {Galluccio}, {Guerrier}, {Heiter}, {Masana},
  {Messineo}, {Mowlavi}, {Nicolas}, {Nienartowicz}, {Pailler}, {Panuzzo},
  {Riclet}, {Roux}, {Seabroke}, {Sordo}, {Th{\'e}venin}, {Gracia-Abril},
  {Portell}, {Teyssier}, {Altmann}, {Andrae}, {Bellas-Velidis}, {Benson},
  {Berthier}, {Blomme}, {Burgess}, {Busonero}, {Busso}, {C{\'a}novas}, {Carry},
  {Cellino}, {Cheek}, {Clementini}, {Damerdji}, {Davidson}, {de Teodoro},
  {Nu{\~n}ez Campos}, {Delchambre}, {Dell'Oro}, {Esquej},
  {Fern{\'a}ndez-Hern{\'a}ndez}, {Fraile}, {Garabato}, {Garc{\'\i}a-Lario},
  {Gosset}, {Haigron}, {Halbwachs}, {Hambly}, {Harrison}, {Hern{\'a}ndez},
  {Hestroffer}, {Hilger}, {Hodgkin}, {Jan{\ss}en}, {Jevardat de Fombelle},
  {Jordan}, {Krone-Martins}, {Lanzafame}, {L{\"o}ffler}, {Marchal}, {Marrese},
  {Moitinho}, {Muinonen}, {Osborne}, {Pancino}, {Pauwels}, {Recio-Blanco},
  {Reyl{\'e}}, {Riello}, {Roegiers}, {Rybizki}, {Sarro}, {Siopis}, {Smith},
  {Sozzetti}, {Utrilla}, {van Leeuwen}, {Abbas}, {{\'A}brah{\'a}m}, {Abreu
  Aramburu}, {Aguado}, {Ajaj}, {Aldea-Montero}, {Altavilla}, {{\'A}lvarez},
  {Alves}, {Anders}, {Anderson}, {Anglada Varela}, {Antoja}, {Baines}, {Baker},
  {Balaguer-N{\'u}{\~n}ez}, {Balbinot}, {Balog}, {Barache}, {Barbato},
  {Barros}, {Barstow}, {Bartolom{\'e}}, {Bassilana}, {Bauchet}, {Becciani},
  {Bellazzini}, {Berihuete}, {Bernet}, {Bertone}, {Bianchi}, {Binnenfeld},
  {Blanco-Cuaresma}, {Boch}, {Bombrun}, {Bossini}, {Bouquillon}, {Bragaglia},
  {Bramante}, {Breedt}, {Bressan}, {Brouillet}, {Brugaletta}, {Bucciarelli},
  {Burlacu}, {Butkevich}, {Buzzi}, {Caffau}, {Cancelliere}, {Cantat-Gaudin},
  {Carballo}, {Carlucci}, {Carnerero}, {Carrasco}, {Casamiquela}, {Castellani},
  {Castro-Ginard}, {Chaoul}, {Charlot}, {Chemin}, {Chiaramida}, {Chiavassa},
  {Chornay}, {Comoretto}, {Contursi}, {Cooper}, {Cornez}, {Cowell}, {Crifo},
  {Cropper}, {Crosta}, {Crowley}, {Dafonte}, {Dapergolas}, {David}, {de
  Laverny}, {De Luise}, {De March}, {de Souza}, {de Torres}, {del Peloso}, {del
  Pozo}, {Delbo}, {Delgado}, {Delisle}, {Demouchy}, {Dharmawardena}, {Diakite},
  {Diener}, {Distefano}, {Dolding}, {Enke}, {Fabre}, {Fabrizio}, {Faigler},
  {Fedorets}, {Fernique}, {Figueras}, {Fournier}, {Fouron}, {Fragkoudi}, {Gai},
  {Garcia-Gutierrez}, {Garcia-Reinaldos}, {Garc{\'\i}a-Torres}, {Garofalo},
  {Gavel}, {Gavras}, {Gerlach}, {Geyer}, {Giacobbe}, {Gilmore}, {Girona},
  {Giuffrida}, {Gomel}, {Gomez}, {Gonz{\'a}lez-N{\'u}{\~n}ez},
  {Gonz{\'a}lez-Santamar{\'\i}a}, {Gonz{\'a}lez-Vidal}, {Granvik}, {Guillout},
  {Guiraud}, {Guti{\'e}rrez-S{\'a}nchez}, {Guy}, {Hatzidimitriou}, {Hauser},
  {Haywood}, {Helmer}, {Helmi}, {Sarmiento}, {Hidalgo}, {H{\l}adczuk}, {Hobbs},
  {Holland}, {Huckle}, {Jardine}, {Jasniewicz}, {Jean-Antoine Piccolo},
  {Jim{\'e}nez-Arranz}, {Juaristi Campillo}, {Julbe}, {Karbevska}, {Kervella},
  {Khanna}, {Kordopatis}, {Korn}, {K{\'o}sp{\'a}l}, {Kostrzewa-Rutkowska},
  {Kruszy{\'n}ska}, {Kun}, {Laizeau}, {Lambert}, {Lanza}, {Lasne}, {Le
  Campion}, {Lebreton}, {Lebzelter}, {Leccia}, {Leclerc}, {Lecoeur-Taibi},
  {Liao}, {Licata}, {Lindstr{\o}m}, {Lister}, {Livanou}, {Lobel}, {Lorca},
  {Loup}, {Madrero Pardo}, {Magdaleno Romeo}, {Managau}, {Mann}, {Manteiga},
  {Marchant}, {Marconi}, {Marcos}, {Marcos Santos}, {Mar{\'\i}n Pina},
  {Marinoni}, {Marocco}, {Marshall}, {Martin Polo}, {Mart{\'\i}n-Fleitas},
  {Marton}, {Mary}, {Masip}, {Massari}, {Mastrobuono-Battisti}, {Mazeh},
  {McMillan}, {Messina}, {Michalik}, {Millar}, {Mints}, {Molina}, {Molinaro},
  {Moln{\'a}r}, {Monari}, {Mongui{\'o}}, {Montegriffo}, {Montero}, {Mor},
  {Mora}, {Morbidelli}, {Morel}, {Morris}, {Muraveva}, {Murphy}, {Musella},
  {Nagy}, {Noval}, {Oca{\~n}a}, {Ogden}, {Ordenovic}, {Osinde}, {Pagani},
  {Pagano}, {Palicio}, {Pallas-Quintela}, {Panahi}, {Payne-Wardenaar},
  {Pe{\~n}alosa Esteller}, {Penttil{\"a}}, {Pichon}, {Piersimoni}, {Pineau},
  {Plachy}, {Plum}, {Poggio}, {Pr{\v{s}}a}, {Pulone}, {Racero}, {Ragaini},
  {Rainer}, {Raiteri}, {Ramos}, {Ramos-Lerate}, {Re Fiorentin}, {Regibo},
  {Richards}, {Rios Diaz}, {Riva}, {Rix}, {Rixon}, {Robichon}, {Robin},
  {Robin}, {Roelens}, {Rogues}, {Rohrbasser}, {Romero-G{\'o}mez}, {Rowell},
  {Royer}, {Ruz Mieres}, {Rybicki}, {Sadowski}, {S{\'a}ez N{\'u}{\~n}ez},
  {Sagrist{\`a} Sell{\'e}s}, {Sahlmann}, {Salguero}, {Samaras}, {Sanchez
  Gimenez}, {Sanna}, {Santove{\~n}a}, {Sarasso}, {Schultheis}, {Sciacca},
  {Segol}, {Segovia}, {S{\'e}gransan}, {Semeux}, {Shahaf}, {Siddiqui},
  {Siebert}, {Siltala}, {Silvelo}, {Slezak}, {Slezak}, {Smart}, {Snaith},
  {Solano}, {Solitro}, {Souami}, {Souchay}, {Spagna}, {Spina}, {Spoto},
  {Steele}, {Steidelm{\"u}ller}, {Stephenson}, {S{\"u}veges}, {Surdej},
  {Szabados}, {Szegedi-Elek}, {Taris}, {Taylor}, {Teixeira}, {Tolomei},
  {Tonello}, {Torra}, {Torra}, {Torralba Elipe}, {Trabucchi}, {Tsounis},
  {Turon}, {Ulla}, {Unger}, {Vaillant}, {vanDillen}, {van Reeven}, {Vanel},
  {Vecchiato}, {Viala}, {Vicente}, {Voutsinas}, {Weiler}, {Wevers},
  {Wyrzykowski}, {Yoldas}, {Yvard}, {Zhao}, {Zorec}, {Zucker}, \&
  {Zwitter}}]{DeRidderEtal2023}
{Gaia Collaboration}, {De Ridder}, J., {Ripepi}, V., {et~al.}
  2023{\natexlab{a}}, \aap, 674, A36

\bibitem[{{Gaia Collaboration} {et~al.}(2019){Gaia Collaboration}, {Eyer},
  {Rimoldini}, {Audard}, {Anderson}, {Nienartowicz}, {Glass}, {Marchal},
  {Grenon}, {Mowlavi}, {Holl}, {Clementini}, {Aerts}, {Mazeh}, {Evans},
  {Szabados}, {Brown}, {Vallenari}, {Prusti}, {de Bruijne}, {Babusiaux},
  {Bailer-Jones}, {Biermann}, {Jansen}, {Jordi}, {Klioner}, {Lammers},
  {Lindegren}, {Luri}, {Mignard}, {Panem}, {Pourbaix}, {Randich}, {Sartoretti},
  {Siddiqui}, {Soubiran}, {van Leeuwen}, {Walton}, {Arenou}, {Bastian},
  {Cropper}, {Drimmel}, {Katz}, {Lattanzi}, {Bakker}, {Cacciari},
  {Casta{\~n}eda}, {Chaoul}, {Cheek}, {De Angeli}, {Fabricius}, {Guerra},
  {Masana}, {Messineo}, {Panuzzo}, {Portell}, {Riello}, {Seabroke}, {Tanga},
  {Th{\'e}venin}, {Gracia-Abril}, {Comoretto}, {Garcia-Reinaldos}, {Teyssier},
  {Altmann}, {Andrae}, {Bellas-Velidis}, {Benson}, {Berthier}, {Blomme},
  {Burgess}, {Busso}, {Carry}, {Cellino}, {Clotet}, {Creevey}, {Davidson}, {De
  Ridder}, {Delchambre}, {Dell'Oro}, {Ducourant},
  {Fern{\'a}ndez-Hern{\'a}ndez}, {Fouesneau}, {Fr{\'e}mat}, {Galluccio},
  {Garc{\'\i}a-Torres}, {Gonz{\'a}lez-N{\'u}{\~n}ez}, {Gonz{\'a}lez-Vidal},
  {Gosset}, {Guy}, {Halbwachs}, {Hambly}, {Harrison}, {Hern{\'a}ndez},
  {Hestroffer}, {Hodgkin}, {Hutton}, {Jasniewicz}, {Jean-Antoine-Piccolo},
  {Jordan}, {Korn}, {Krone-Martins}, {Lanzafame}, {Lebzelter}, {L{\"o}ffler},
  {Manteiga}, {Marrese}, {Mart{\'\i}n-Fleitas}, {Moitinho}, {Mora}, {Muinonen},
  {Osinde}, {Pancino}, {Pauwels}, {Petit}, {Recio-Blanco}, {Richards}, {Robin},
  {Sarro}, {Siopis}, {Smith}, {Sozzetti}, {S{\"u}veges}, {Torra}, {van Reeven},
  {Abbas}, {Abreu Aramburu}, {Accart}, {Altavilla}, {{\'A}lvarez}, {Alvarez},
  {Alves}, {Andrei}, {Anglada Varela}, {Antiche}, {Antoja}, {Arcay},
  {Astraatmadja}, {Bach}, {Baker}, {Balaguer-N{\'u}{\~n}ez}, {Balm}, {Barache},
  {Barata}, {Barbato}, {Barblan}, {Barklem}, {Barrado}, {Barros}, {Barstow},
  {Bartholom{\'e} Mu{\~n}oz}, {Bassilana}, {Becciani}, {Bellazzini},
  {Berihuete}, {Bertone}, {Bianchi}, {Bienaym{\'e}}, {Blanco-Cuaresma}, {Boch},
  {Boeche}, {Bombrun}, {Borrachero}, {Bossini}, {Bouquillon}, {Bourda},
  {Bragaglia}, {Bramante}, {Breddels}, {Bressan}, {Brouillet},
  {Br{\"u}semeister}, {Brugaletta}, {Bucciarelli}, {Burlacu}, {Busonero},
  {Butkevich}, {Buzzi}, {Caffau}, {Cancelliere}, {Cannizzaro}, {Cantat-Gaudin},
  {Carballo}, {Carlucci}, {Carrasco}, {Casamiquela}, {Castellani},
  {Castro-Ginard}, {Charlot}, {Chemin}, {Chiavassa}, {Cocozza}, {Costigan},
  {Cowell}, {Crifo}, {Crosta}, {Crowley}, {Cuypers}, {Dafonte}, {Damerdji},
  {Dapergolas}, {David}, {David}, {de Laverny}, {De Luise}, {De March}, {de
  Martino}, {de Souza}, {de Torres}, {Debosscher}, {del Pozo}, {Delbo},
  {Delgado}, {Delgado}, {Diakite}, {Diener}, {Distefano}, {Dolding},
  {Drazinos}, {Dur{\'a}n}, {Edvardsson}, {Enke}, {Eriksson}, {Esquej}, {Eynard
  Bontemps}, {Fabre}, {Fabrizio}, {Faigler}, {Falc{\~a}o}, {Farr{\`a}s Casas},
  {Federici}, {Fedorets}, {Fernique}, {Figueras}, {Filippi}, {Findeisen},
  {Fonti}, {Fraile}, {Fraser}, {Fr{\'e}zouls}, {Gai}, {Galleti}, {Garabato},
  {Garc{\'\i}a-Sedano}, {Garofalo}, {Garralda}, {Gavel}, {Gavras}, {Gerssen},
  {Geyer}, {Giacobbe}, {Gilmore}, {Girona}, {Giuffrida}, {Gomes}, {Granvik},
  {Gueguen}, {Guerrier}, {Guiraud}, {Guti{\'e}rrez-S{\'a}nchez}, {Haigron},
  {Hatzidimitriou}, {Hauser}, {Haywood}, {Heiter}, {Helmi}, {Heu}, {Hilger},
  {Hobbs}, {Hofmann}, {Holland}, {Huckle}, {Hypki}, {Icardi}, {Jan{\ss}en},
  {Jevardat de Fombelle}, {Jonker}, {Juh{\'a}sz}, {Julbe}, {Karampelas},
  {Kewley}, {Klar}, {Kochoska}, {Kohley}, {Kolenberg}, {Kontizas}, {Kontizas},
  {Koposov}, {Kordopatis}, {Kostrzewa-Rutkowska}, {Koubsky}, {Lambert},
  {Lanza}, {Lasne}, {Lavigne}, {Le Fustec}, {Le Poncin-Lafitte}, {Lebreton},
  {Leccia}, {Leclerc}, {Lecoeur-Taibi}, {Lenhardt}, {Leroux}, {Liao}, {Licata},
  {Lindstr{\o}m}, {Lister}, {Livanou}, {Lobel}, {L{\'o}pez}, {Lorenz},
  {Managau}, {Mann}, {Mantelet}, {Marchant}, {Marconi}, {Marinoni},
  {Marschalk{\'o}}, {Marshall}, {Martino}, {Marton}, {Mary}, {Massari},
  {Matijevi{\v{c}}}, {McMillan}, {Messina}, {Michalik}, {Millar}, {Molina},
  {Molinaro}, {Moln{\'a}r}, {Montegriffo}, {Mor}, {Morbidelli}, {Morel},
  {Morgenthaler}, {Morris}, {Mulone}, {Muraveva}, {Musella}, {Nelemans},
  {Nicastro}, {Noval}, {O'Mullane}, {Ord{\'e}novic}, {Ord{\'o}{\~n}ez-Blanco},
  {Osborne}, {Pagani}, {Pagano}, {Pailler}, {Palacin}, {Palaversa}, {Panahi},
  {Pawlak}, {Piersimoni}, {Pineau}, {Plachy}, {Plum}, {Poggio}, {Poujoulet},
  {Pr{\v{s}}a}, {Pulone}, {Racero}, {Ragaini}, {Rambaux}, {Ramos-Lerate},
  {Regibo}, {Reyl{\'e}}, {Riclet}, {Ripepi}, {Riva}, {Rivard}, {Rixon},
  {Roegiers}, {Roelens}, {Romero-G{\'o}mez}, {Rowell}, {Royer}, {Ruiz-Dern},
  {Sadowski}, {Sagrist{\`a} Sell{\'e}s}, {Sahlmann}, {Salgado}, {Salguero},
  {Sanna}, {Santana-Ros}, {Sarasso}, {Savietto}, {Schultheis}, {Sciacca},
  {Segol}, {Segovia}, {S{\'e}gransan}, {Shih}, {Siltala}, {Silva}, {Smart},
  {Smith}, {Solano}, {Solitro}, {Sordo}, {Soria Nieto}, {Souchay}, {Spagna},
  {Spoto}, {Stampa}, {Steele}, {Steidelm{\"u}ller}, {Stephenson}, {Stoev},
  {Suess}, {Surdej}, {Szegedi-Elek}, {Tapiador}, {Taris}, {Tauran}, {Taylor},
  {Teixeira}, {Terrett}, {Teyssandier}, {Thuillot}, {Titarenko}, {Torra
  Clotet}, {Turon}, {Ulla}, {Utrilla}, {Uzzi}, {Vaillant}, {Valentini},
  {Valette}, {van Elteren}, {Van Hemelryck}, {van Leeuwen}, {Vaschetto},
  {Vecchiato}, {Veljanoski}, {Viala}, {Vicente}, {Vogt}, {von Essen}, {Voss},
  {Votruba}, {Voutsinas}, {Walmsley}, {Weiler}, {Wertz}, {Wevers},
  {Wyrzykowski}, {Yoldas}, {{\v{Z}}erjal}, {Ziaeepour}, {Zorec}, {Zschocke},
  {Zucker}, {Zurbach}, \& {Zwitter}}]{EyerEtal2019}
{Gaia Collaboration}, {Eyer}, L., {Rimoldini}, L., {et~al.} 2019, \aap, 623,
  A110

\bibitem[{{Gaia Collaboration} {et~al.}(2024){Gaia Collaboration}, {Panuzzo},
  {Mazeh}, {Arenou}, {Holl}, {Caffau}, {Jorissen}, {Babusiaux}, {Gavras},
  {Sahlmann}, {Bastian}, {Wyrzykowski}, {Eyer}, {Leclerc}, {Bauchet},
  {Bombrun}, {Mowlavi}, {Seabroke}, {Teyssier}, {Balbinot}, {Helmi}, {Brown},
  {Vallenari}, {Prusti}, {de Bruijne}, {Barbier}, {Biermann}, {Creevey},
  {Ducourant}, {Evans}, {Guerra}, {Hutton}, {Jordi}, {Klioner}, {Lammers},
  {Lindegren}, {Luri}, {Mignard}, {Nicolas}, {Randich}, {Sartoretti},
  {Smiljanic}, {Tanga}, {Walton}, {Aerts}, {Bailer-Jones}, {Cropper},
  {Drimmel}, {Jansen}, {Katz}, {Lattanzi}, {Soubiran}, {Th{\'e}venin}, {van
  Leeuwen}, {Andrae}, {Audard}, {Bakker}, {Blomme}, {Casta{\~n}eda}, {De
  Angeli}, {Fabricius}, {Fouesneau}, {Fr{\'e}mat}, {Galluccio}, {Guerrier},
  {Heiter}, {Masana}, {Messineo}, {Nienartowicz}, {Pailler}, {Riclet}, {Roux},
  {Sordo}, {Gracia-Abril}, {Portell}, {Altmann}, {Benson}, {Berthier},
  {Burgess}, {Busonero}, {Busso}, {Cacciari}, {C{\'a}novas}, {Carrasco},
  {Carry}, {Cellino}, {Cheek}, {Clementini}, {Damerdji}, {Davidson}, {de
  Teodoro}, {Delchambre}, {Dell'Oro}, {Fraile Garcia}, {Garabato},
  {Garc{\'\i}a-Lario}, {Haigron}, {Hambly}, {Harrison}, {Hatzidimitriou},
  {Hern{\'a}ndez}, {Hestroffer}, {Hodgkin}, {Jamal}, {Jevardat de Fombelle},
  {Jordan}, {Krone-Martins}, {Lanzafame}, {L{\"o}ffler}, {Lorca}, {Marchal},
  {Marrese}, {Moitinho}, {Muinonen}, {Nu{\~n}ez Campos}, {Oreshina-Slezak},
  {Osborne}, {Pancino}, {Pauwels}, {Recio-Blanco}, {Riello}, {Rimoldini},
  {Robin}, {Roegiers}, {Sarro}, {Schultheis}, {Smith}, {Sozzetti}, {Utrilla},
  {van Leeuwen}, {Weingrill}, {Abbas}, {{\'A}brah{\'a}m}, {Abreu Aramburu},
  {Ahmed}, {Altavilla}, {{\'A}lvarez}, {Anders}, {Anderson}, {Anglada Varela},
  {Antoja}, {Baig}, {Baines}, {Baker}, {Balaguer-N{\'u}{\~n}ez}, {Balog},
  {Barache}, {Barros}, {Barstow}, {Bartolom{\'e}}, {Bashi}, {Bassilana},
  {Baudeau}, {Becciani}, {Bedin}, {Bellas-Velidis}, {Bellazzini}, {Beordo},
  {Bernet}, {Bertolotto}, {Bertone}, {Bianchi}, {Binnenfeld},
  {Blanco-Cuaresma}, {Bland-Hawthorn}, {Blazere}, {Boch}, {Bossini},
  {Bouquillon}, {Bragaglia}, {Braine}, {Bratsolis}, {Breedt}, {Bressan},
  {Brouillet}, {Brugaletta}, {Bucciarelli}, {Butkevich}, {Buzzi}, {Camut},
  {Cancelliere}, {Cantat-Gaudin}, {Capilla Guilarte}, {Carballo}, {Carlucci},
  {Carnerero}, {Carretero}, {Carton}, {Casamiquela}, {Casey}, {Castellani},
  {Castro-Ginard}, {Ceraj}, {Cesare}, {Charlot}, {Chaudet}, {Chemin},
  {Chiavassa}, {Chornay}, {Chosson}, {Cooper}, {Cornez}, {Cowell}, {Crosta},
  {Crowley}, {Cruz Reyes}, {Dafonte}, {Dal Ponte}, {David}, {de Laverny}, {De
  Luise}, {De March}, {de Torres}, {del Peloso}, {Delbo}, {Delgado}, {Delisle},
  {Demouchy}, {Denis}, {Dharmawardena}, {Di Giacomo}, {Diener}, {Distefano},
  {Dolding}, {Dsilva}, {Enke}, {Fabre}, {Fabrizio}, {Faigler}, {Fatovi{\'c}},
  {Fedorets}, {Fern{\'a}ndez-Hern{\'a}ndez}, {Fernique}, {Figueras}, {Fouron},
  {Fragkoudi}, {Gai}, {Galinier}, {Garcia-Serrano}, {Garc{\'\i}a-Torres},
  {Garofalo}, {Gerlach}, {Geyer}, {Giacobbe}, {Gilmore}, {Girona}, {Giuffrida},
  {Gomboc}, {Gomez}, {Gonz{\'a}lez-Santamar{\'\i}a}, {Gosset}, {Granvik},
  {Gregori Barrera}, {Guti{\'e}rrez-S{\'a}nchez}, {Haywood}, {Helmer},
  {Hidalgo}, {Hilger}, {Hobbs}, {Hottier}, {Huckle}, {Jim{\'e}nez-Arranz},
  {Juaristi Campillo}, {Kaczmarek}, {Kervella}, {Khanna}, {Kontizas},
  {Kordopatis}, {Korn}, {K{\'o}sp{\'a}l}, {Kostrzewa-Rutkowska},
  {Kruszy{\'n}ska}, {Kun}, {Lambert}, {Lanza}, {Lebreton}, {Lebzelter},
  {Leccia}, {Lecoutre}, {Liao}, {Liberato}, {Licata}, {Livanou}, {Lobel},
  {L{\'o}pez-Miralles}, {Loup}, {Madar{\'a}sz}, {Mahy}, {Mann}, {Manteiga},
  {Marcellino}, {Marchant}, {Marconi}, {Mar{\'\i}n Pina}, {Marinoni},
  {Marshall}, {Mart{\'\i}n Lozano}, {Martin Polo}, {Mart{\'\i}n-Fleitas},
  {Marton}, {Mascarenhas}, {Masip}, {Mastrobuono-Battisti}, {McMillan},
  {Meichsner}, {Merc}, {Messina}, {Millar}, {Mints}, {Mohamed}, {Molina},
  {Molinaro}, {Moln{\'a}r}, {Mongui{\'o}}, {Montegriffo}, {Monti}, {Mora},
  {Morbidelli}, {Morris}, {Mudimadugula}, {Muraveva}, {Musella}, {Nagy},
  {Nardetto}, {Navarrete}, {Oh}, {Ordenovic}, {Orenstein}, {Pagani}, {Pagano},
  {Palaversa}, {Palicio}, {Pallas-Quintela}, {Pawlak}, {Penttil{\"a}},
  {Pesciullesi}, {Pinamonti}, {Plachy}, {Planquart}, {Plum}, {Poggio},
  {Pourbaix}, {Price-Whelan}, {Pulone}, {Rabin}, {Rainer}, {Raiteri}, {Ramos},
  {Ramos-Lerate}, {Ratajczak}, {Re Fiorentin}, {Regibo}, {Reyl{\'e}}, {Ripepi},
  {Riva}, {Rix}, {Rixon}, {Robert}, {Robichon}, {Robin}, {Romero-G{\'o}mez},
  {Rowell}, {Ruz Mieres}, {Rybicki}, {Sadowski}, {Sagrist{\`a} Sell{\'e}s},
  {Sanna}, {Santove{\~n}a}, {Sarasso}, {Sarmiento}, {Sarrate Riera}, {Sciacca},
  {S{\'e}gransan}, {Semczuk}, {Shahaf}, {Siebert}, {Slezak}, {Smart}, {Snaith},
  {Solano}, {Solitro}, {Souami}, {Souchay}, {Spitoni}, {Spoto}, {Squillante},
  {Steele}, {Steidelm{\"u}ller}, {Surdej}, {Szabados}, {Taris}, {Taylor},
  {Teixeira}, {Tepper-Garcia}, {Thuillot}, {Tolomei}, {Tonello}, {Torra},
  {Torralba Elipe}, {Trabucchi}, {Trentin}, {Tsantaki}, {Turon}, {Ulla},
  {Unger}, {Valtchanov}, {Vanel}, {Vecchiato}, {Vicente}, {Villar}, {Weiler},
  {Zhao}, {Zorec}, {Zucker}, {{\v{Z}}upi{\'c}}, \& {Zwitter}}]{PanuzzoEtal2024}
{Gaia Collaboration}, {Panuzzo}, P., {Mazeh}, T., {et~al.} 2024, \aap, 686, L2

\bibitem[{{Gaia Collaboration} {et~al.}(2016){Gaia Collaboration}, {Prusti},
  {de Bruijne}, {Brown}, {Vallenari}, {Babusiaux}, {Bailer-Jones}, {Bastian},
  {Biermann}, {Evans}, {Eyer}, {Jansen}, {Jordi}, {Klioner}, {Lammers},
  {Lindegren}, {Luri}, {Mignard}, {Milligan}, {Panem}, {Poinsignon},
  {Pourbaix}, {Randich}, {Sarri}, {Sartoretti}, {Siddiqui}, {Soubiran},
  {Valette}, {van Leeuwen}, {Walton}, {Aerts}, {Arenou}, {Cropper}, {Drimmel},
  {H{\o}g}, {Katz}, {Lattanzi}, {O'Mullane}, {Grebel}, {Holland}, {Huc},
  {Passot}, {Bramante}, {Cacciari}, {Casta{\~n}eda}, {Chaoul}, {Cheek}, {De
  Angeli}, {Fabricius}, {Guerra}, {Hern{\'a}ndez}, {Jean-Antoine-Piccolo},
  {Masana}, {Messineo}, {Mowlavi}, {Nienartowicz}, {Ord{\'o}{\~n}ez-Blanco},
  {Panuzzo}, {Portell}, {Richards}, {Riello}, {Seabroke}, {Tanga},
  {Th{\'e}venin}, {Torra}, {Els}, {Gracia-Abril}, {Comoretto},
  {Garcia-Reinaldos}, {Lock}, {Mercier}, {Altmann}, {Andrae}, {Astraatmadja},
  {Bellas-Velidis}, {Benson}, {Berthier}, {Blomme}, {Busso}, {Carry},
  {Cellino}, {Clementini}, {Cowell}, {Creevey}, {Cuypers}, {Davidson}, {De
  Ridder}, {de Torres}, {Delchambre}, {Dell'Oro}, {Ducourant}, {Fr{\'e}mat},
  {Garc{\'\i}a-Torres}, {Gosset}, {Halbwachs}, {Hambly}, {Harrison}, {Hauser},
  {Hestroffer}, {Hodgkin}, {Huckle}, {Hutton}, {Jasniewicz}, {Jordan},
  {Kontizas}, {Korn}, {Lanzafame}, {Manteiga}, {Moitinho}, {Muinonen},
  {Osinde}, {Pancino}, {Pauwels}, {Petit}, {Recio-Blanco}, {Robin}, {Sarro},
  {Siopis}, {Smith}, {Smith}, {Sozzetti}, {Thuillot}, {van Reeven}, {Viala},
  {Abbas}, {Abreu Aramburu}, {Accart}, {Aguado}, {Allan}, {Allasia},
  {Altavilla}, {{\'A}lvarez}, {Alves}, {Anderson}, {Andrei}, {Anglada Varela},
  {Antiche}, {Antoja}, {Ant{\'o}n}, {Arcay}, {Atzei}, {Ayache}, {Bach},
  {Baker}, {Balaguer-N{\'u}{\~n}ez}, {Barache}, {Barata}, {Barbier}, {Barblan},
  {Baroni}, {Barrado y Navascu{\'e}s}, {Barros}, {Barstow}, {Becciani},
  {Bellazzini}, {Bellei}, {Bello Garc{\'\i}a}, {Belokurov}, {Bendjoya},
  {Berihuete}, {Bianchi}, {Bienaym{\'e}}, {Billebaud}, {Blagorodnova},
  {Blanco-Cuaresma}, {Boch}, {Bombrun}, {Borrachero}, {Bouquillon}, {Bourda},
  {Bouy}, {Bragaglia}, {Breddels}, {Brouillet}, {Br{\"u}semeister},
  {Bucciarelli}, {Budnik}, {Burgess}, {Burgon}, {Burlacu}, {Busonero}, {Buzzi},
  {Caffau}, {Cambras}, {Campbell}, {Cancelliere}, {Cantat-Gaudin}, {Carlucci},
  {Carrasco}, {Castellani}, {Charlot}, {Charnas}, {Charvet}, {Chassat},
  {Chiavassa}, {Clotet}, {Cocozza}, {Collins}, {Collins}, {Costigan}, {Crifo},
  {Cross}, {Crosta}, {Crowley}, {Dafonte}, {Damerdji}, {Dapergolas}, {David},
  {David}, {De Cat}, {de Felice}, {de Laverny}, {De Luise}, {De March}, {de
  Martino}, {de Souza}, {Debosscher}, {del Pozo}, {Delbo}, {Delgado},
  {Delgado}, {di Marco}, {Di Matteo}, {Diakite}, {Distefano}, {Dolding}, {Dos
  Anjos}, {Drazinos}, {Dur{\'a}n}, {Dzigan}, {Ecale}, {Edvardsson}, {Enke},
  {Erdmann}, {Escolar}, {Espina}, {Evans}, {Eynard Bontemps}, {Fabre},
  {Fabrizio}, {Faigler}, {Falc{\~a}o}, {Farr{\`a}s Casas}, {Faye}, {Federici},
  {Fedorets}, {Fern{\'a}ndez-Hern{\'a}ndez}, {Fernique}, {Fienga}, {Figueras},
  {Filippi}, {Findeisen}, {Fonti}, {Fouesneau}, {Fraile}, {Fraser}, {Fuchs},
  {Furnell}, {Gai}, {Galleti}, {Galluccio}, {Garabato}, {Garc{\'\i}a-Sedano},
  {Gar{\'e}}, {Garofalo}, {Garralda}, {Gavras}, {Gerssen}, {Geyer}, {Gilmore},
  {Girona}, {Giuffrida}, {Gomes}, {Gonz{\'a}lez-Marcos},
  {Gonz{\'a}lez-N{\'u}{\~n}ez}, {Gonz{\'a}lez-Vidal}, {Granvik}, {Guerrier},
  {Guillout}, {Guiraud}, {G{\'u}rpide}, {Guti{\'e}rrez-S{\'a}nchez}, {Guy},
  {Haigron}, {Hatzidimitriou}, {Haywood}, {Heiter}, {Helmi}, {Hobbs},
  {Hofmann}, {Holl}, {Holland}, {Hunt}, {Hypki}, {Icardi}, {Irwin}, {Jevardat
  de Fombelle}, {Jofr{\'e}}, {Jonker}, {Jorissen}, {Julbe}, {Karampelas},
  {Kochoska}, {Kohley}, {Kolenberg}, {Kontizas}, {Koposov}, {Kordopatis},
  {Koubsky}, {Kowalczyk}, {Krone-Martins}, {Kudryashova}, {Kull}, {Bachchan},
  {Lacoste-Seris}, {Lanza}, {Lavigne}, {Le Poncin-Lafitte}, {Lebreton},
  {Lebzelter}, {Leccia}, {Leclerc}, {Lecoeur-Taibi}, {Lemaitre}, {Lenhardt},
  {Leroux}, {Liao}, {Licata}, {Lindstr{\o}m}, {Lister}, {Livanou}, {Lobel},
  {L{\"o}ffler}, {L{\'o}pez}, {Lopez-Lozano}, {Lorenz}, {Loureiro},
  {MacDonald}, {Magalh{\~a}es Fernandes}, {Managau}, {Mann}, {Mantelet},
  {Marchal}, {Marchant}, {Marconi}, {Marie}, {Marinoni}, {Marrese},
  {Marschalk{\'o}}, {Marshall}, {Mart{\'\i}n-Fleitas}, {Martino}, {Mary},
  {Matijevi{\v{c}}}, {Mazeh}, {McMillan}, {Messina}, {Mestre}, {Michalik},
  {Millar}, {Miranda}, {Molina}, {Molinaro}, {Molinaro}, {Moln{\'a}r},
  {Moniez}, {Montegriffo}, {Monteiro}, {Mor}, {Mora}, {Morbidelli}, {Morel},
  {Morgenthaler}, {Morley}, {Morris}, {Mulone}, {Muraveva}, {Musella},
  {Narbonne}, {Nelemans}, {Nicastro}, {Noval}, {Ord{\'e}novic},
  {Ordieres-Mer{\'e}}, {Osborne}, {Pagani}, {Pagano}, {Pailler}, {Palacin},
  {Palaversa}, {Parsons}, {Paulsen}, {Pecoraro}, {Pedrosa}, {Pentik{\"a}inen},
  {Pereira}, {Pichon}, {Piersimoni}, {Pineau}, {Plachy}, {Plum}, {Poujoulet},
  {Pr{\v{s}}a}, {Pulone}, {Ragaini}, {Rago}, {Rambaux}, {Ramos-Lerate},
  {Ranalli}, {Rauw}, {Read}, {Regibo}, {Renk}, {Reyl{\'e}}, {Ribeiro},
  {Rimoldini}, {Ripepi}, {Riva}, {Rixon}, {Roelens}, {Romero-G{\'o}mez},
  {Rowell}, {Royer}, {Rudolph}, {Ruiz-Dern}, {Sadowski}, {Sagrist{\`a}
  Sell{\'e}s}, {Sahlmann}, {Salgado}, {Salguero}, {Sarasso}, {Savietto},
  {Schnorhk}, {Schultheis}, {Sciacca}, {Segol}, {Segovia}, {Segransan},
  {Serpell}, {Shih}, {Smareglia}, {Smart}, {Smith}, {Solano}, {Solitro},
  {Sordo}, {Soria Nieto}, {Souchay}, {Spagna}, {Spoto}, {Stampa}, {Steele},
  {Steidelm{\"u}ller}, {Stephenson}, {Stoev}, {Suess}, {S{\"u}veges}, {Surdej},
  {Szabados}, {Szegedi-Elek}, {Tapiador}, {Taris}, {Tauran}, {Taylor},
  {Teixeira}, {Terrett}, {Tingley}, {Trager}, {Turon}, {Ulla}, {Utrilla},
  {Valentini}, {van Elteren}, {Van Hemelryck}, {van Leeuwen}, {Varadi},
  {Vecchiato}, {Veljanoski}, {Via}, {Vicente}, {Vogt}, {Voss}, {Votruba},
  {Voutsinas}, {Walmsley}, {Weiler}, {Weingrill}, {Werner}, {Wevers},
  {Whitehead}, {Wyrzykowski}, {Yoldas}, {{\v{Z}}erjal}, {Zucker}, {Zurbach},
  {Zwitter}, {Alecu}, {Allen}, {Allende Prieto}, {Amorim},
  {Anglada-Escud{\'e}}, {Arsenijevic}, {Azaz}, {Balm}, {Beck}, {Bernstein},
  {Bigot}, {Bijaoui}, {Blasco}, {Bonfigli}, {Bono}, {Boudreault}, {Bressan},
  {Brown}, {Brunet}, {Bunclark}, {Buonanno}, {Butkevich}, {Carret}, {Carrion},
  {Chemin}, {Ch{\'e}reau}, {Corcione}, {Darmigny}, {de Boer}, {de Teodoro}, {de
  Zeeuw}, {Delle Luche}, {Domingues}, {Dubath}, {Fodor}, {Fr{\'e}zouls},
  {Fries}, {Fustes}, {Fyfe}, {Gallardo}, {Gallegos}, {Gardiol}, {Gebran},
  {Gomboc}, {G{\'o}mez}, {Grux}, {Gueguen}, {Heyrovsky}, {Hoar}, {Iannicola},
  {Isasi Parache}, {Janotto}, {Joliet}, {Jonckheere}, {Keil}, {Kim},
  {Klagyivik}, {Klar}, {Knude}, {Kochukhov}, {Kolka}, {Kos}, {Kutka}, {Lainey},
  {LeBouquin}, {Liu}, {Loreggia}, {Makarov}, {Marseille}, {Martayan},
  {Martinez-Rubi}, {Massart}, {Meynadier}, {Mignot}, {Munari}, {Nguyen},
  {Nordlander}, {Ocvirk}, {O'Flaherty}, {Olias Sanz}, {Ortiz}, {Osorio},
  {Oszkiewicz}, {Ouzounis}, {Palmer}, {Park}, {Pasquato}, {Peltzer}, {Peralta},
  {P{\'e}turaud}, {Pieniluoma}, {Pigozzi}, {Poels}, {Prat}, {Prod'homme},
  {Raison}, {Rebordao}, {Risquez}, {Rocca-Volmerange}, {Rosen}, {Ruiz-Fuertes},
  {Russo}, {Sembay}, {Serraller Vizcaino}, {Short}, {Siebert}, {Silva},
  {Sinachopoulos}, {Slezak}, {Soffel}, {Sosnowska}, {Strai{\v{z}}ys}, {ter
  Linden}, {Terrell}, {Theil}, {Tiede}, {Troisi}, {Tsalmantza}, {Tur},
  {Vaccari}, {Vachier}, {Valles}, {Van Hamme}, {Veltz}, {Virtanen}, {Wallut},
  {Wichmann}, {Wilkinson}, {Ziaeepour}, \& {Zschocke}}]{PrustiEtal2016}
{Gaia Collaboration}, {Prusti}, T., {de Bruijne}, J.~H.~J., {et~al.} 2016,
  \aap, 595, A1

\bibitem[{{Gaia Collaboration} {et~al.}(2023{\natexlab{b}}){Gaia
  Collaboration}, {Trabucchi}, {Mowlavi}, {Lebzelter}, {Lecoeur-Taibi},
  {Audard}, {Eyer}, {Garc{\'\i}a-Lario}, {Gavras}, {Holl}, {Jevardat de
  Fombelle}, {Nienartowicz}, {Rimoldini}, {Sartoretti}, {Blomme}, {Fr{\'e}mat},
  {Marchal}, {Damerdji}, {Brown}, {Guerrier}, {Panuzzo}, {Katz}, {Seabroke},
  {Benson}, {Haigron}, {Smith}, {Lobel}, {Vallenari}, {Prusti}, {de Bruijne},
  {Arenou}, {Babusiaux}, {Barbier}, {Biermann}, {Creevey}, {Ducourant},
  {Evans}, {Guerra}, {Hutton}, {Jordi}, {Klioner}, {Lammers}, {Lindegren},
  {Luri}, {Mignard}, {Randich}, {Smiljanic}, {Tanga}, {Walton}, {Bailer-Jones},
  {Bastian}, {Cropper}, {Drimmel}, {Lattanzi}, {Soubiran}, {van Leeuwen},
  {Bakker}, {Casta{\~n}eda}, {De Angeli}, {Fabricius}, {Fouesneau},
  {Galluccio}, {Masana}, {Messineo}, {Nicolas}, {Pailler}, {Riclet}, {Roux},
  {Sordo}, {Th{\'e}venin}, {Gracia-Abril}, {Portell}, {Teyssier}, {Altmann},
  {Berthier}, {Burgess}, {Busonero}, {Busso}, {C{\'a}novas}, {Carry}, {Cheek},
  {Clementini}, {Davidson}, {de Teodoro}, {Delchambre}, {Dell'Oro}, {Fraile
  Garcia}, {Garabato}, {Garralda Torres}, {Hambly}, {Harrison},
  {Hatzidimitriou}, {Hern{\'a}ndez}, {Hodgkin}, {Jamal}, {Jordan},
  {Krone-Martins}, {Lanzafame}, {L{\"o}ffler}, {Lorca}, {Marrese}, {Moitinho},
  {Muinonen}, {Nu{\~n}ez Campos}, {Oreshina-Slezak}, {Osborne}, {Pancino},
  {Pauwels}, {Recio-Blanco}, {Riello}, {Robin}, {Roegiers}, {Sarro},
  {Schultheis}, {Siopis}, {Sozzetti}, {Utrilla}, {van Leeuwen}, {Weingrill},
  {Abbas}, {{\'A}brah{\'a}m}, {Abreu Aramburu}, {Aerts}, {Altavilla},
  {{\'A}lvarez}, {Alves}, {Anders}, {Anderson}, {Antoja}, {Baines}, {Baker},
  {Balog}, {Barache}, {Barbato}, {Barros}, {Barstow}, {Bartolom{\'e}}, {Bashi},
  {Bauchet}, {Baudeau}, {Becciani}, {Bedin}, {Bellas-Velidis}, {Bellazzini},
  {Beordo}, {Berihuete}, {Bernet}, {Bertolotto}, {Bertone}, {Bianchi},
  {Binnenfeld}, {Blazere}, {Boch}, {Bombrun}, {Bouquillon}, {Bragaglia},
  {Braine}, {Bramante}, {Breedt}, {Bressan}, {Brouillet}, {Brugaletta},
  {Bucciarelli}, {Butkevich}, {Buzzi}, {Caffau}, {Cancelliere}, {Cannizzo},
  {Carballo}, {Carlucci}, {Carnerero}, {Carrasco}, {Carretero}, {Carton},
  {Casamiquela}, {Castellani}, {Castro-Ginard}, {Cesare}, {Charlot}, {Chemin},
  {Chiaramida}, {Chiavassa}, {Chornay}, {Collins}, {Contursi}, {Cooper},
  {Cornez}, {Crosta}, {Crowley}, {Dafonte}, {David}, {de Laverny}, {De Luise},
  {De March}, {De Ridder}, {de Souza}, {de Torres}, {del Peloso}, {Delbo},
  {Delgado}, {Dharmawardena}, {Diakite}, {Diener}, {Distefano}, {Dolding},
  {Dsilva}, {Dur{\'a}n}, {Enke}, {Esquej}, {Fabre}, {Fabrizio}, {Faigler},
  {Fatovi{\'c}}, {Fedorets}, {Fern{\'a}ndez-Hern{\'a}ndez}, {Fernique},
  {Figueras}, {Fournier}, {Fouron}, {Gai}, {Galinier}, {Garcia-Gutierrez},
  {Garc{\'\i}a-Torres}, {Garofalo}, {Gerlach}, {Geyer}, {Giacobbe}, {Gilmore},
  {Girona}, {Giuffrida}, {Gomel}, {Gomez}, {Gonz{\'a}lez-N{\'u}{\~n}ez},
  {Gonz{\'a}lez-Santamar{\'\i}a}, {Gosset}, {Granvik}, {Gregori Barrera},
  {Guti{\'e}rrez-S{\'a}nchez}, {Haywood}, {Helmer}, {Helmi}, {Henares},
  {Hidalgo}, {Hilger}, {Hobbs}, {Hottier}, {Huckle}, {Jab{\l}o{\'n}ska},
  {Jansen}, {Jim{\'e}nez-Arranz}, {Juaristi Campillo}, {Khanna}, {Kordopatis},
  {K{\'o}sp{\'a}l}, {Kostrzewa-Rutkowska}, {Kun}, {Lambert}, {Lanza}, {Le
  Campion}, {Lebreton}, {Leccia}, {Lecoutre}, {Liao}, {Liberato}, {Licata},
  {Lindstr{\o}m}, {Lister}, {Livanou}, {Loup}, {Mahy}, {Mann}, {Manteiga},
  {Marchant}, {Marconi}, {Mar{\'\i}n Pina}, {Marinoni}, {Marshall},
  {Mart{\'\i}n Lozano}, {Mart{\'\i}n-Fleitas}, {Marton}, {Mary}, {Masip},
  {Massari}, {Mastrobuono-Battisti}, {Mazeh}, {McMillan}, {Meichsner},
  {Messina}, {Michalik}, {Millar}, {Mints}, {Molina}, {Molinaro}, {Moln{\'a}r},
  {Monari}, {Mongui{\'o}}, {Montegriffo}, {Montero}, {Mor}, {Mora},
  {Morbidelli}, {Morel}, {Morris}, {Munoz}, {Muraveva}, {Murphy}, {Musella},
  {Nagy}, {Nieto}, {Noval}, {Ogden}, {Ordenovic}, {Pagani}, {Pagano},
  {Palaversa}, {Palicio}, {Pallas-Quintela}, {Panahi}, {Panem},
  {Payne-Wardenaar}, {Pegoraro}, {Penttil{\"a}}, {Pesciullesi}, {Piersimoni},
  {Pinamonti}, {Pineau}, {Plachy}, {Plum}, {Poggio}, {Pourbaix}, {Pr{\v{s}}a},
  {Pulone}, {Racero}, {Rainer}, {Raiteri}, {Ramos}, {Ramos-Lerate},
  {Ratajczak}, {Re Fiorentin}, {Regibo}, {Reyl{\'e}}, {Ripepi}, {Riva}, {Rix},
  {Rixon}, {Robichon}, {Robin}, {Romero-G{\'o}mez}, {Rowell}, {Royer}, {Ruz
  Mieres}, {Rybicki}, {Sadowski}, {S{\'a}ez N{\'u}{\~n}ez}, {Sagrist{\`a}
  Sell{\'e}s}, {Sahlmann}, {Sanchez Gimenez}, {Sanna}, {Santove{\~n}a},
  {Sarasso}, {Sarrate Riera}, {Sciacca}, {Segovia}, {S{\'e}gransan}, {Shahaf},
  {Siebert}, {Siltala}, {Slezak}, {Smart}, {Snaith}, {Solano}, {Solitro},
  {Souami}, {Souchay}, {Spina}, {Spitoni}, {Spoto}, {Squillante}, {Steele},
  {Steidelm{\"u}ller}, {Surdej}, {Szabados}, {Taris}, {Taylor}, {Teixeira},
  {Tisani{\'c}}, {Tolomei}, {Torra}, {Torralba Elipe}, {Tsantaki}, {Ulla},
  {Unger}, {Vanel}, {Vecchiato}, {Vicente}, {Voutsinas}, {Weiler},
  {Wyrzykowski}, {Zhao}, {Zorec}, {Zwitter}, {Balaguer-Nunez}, {Leclerc},
  {Morgenthaler}, {Robert}, \& {Zucker}}]{TrabucchiEtal2023}
{Gaia Collaboration}, {Trabucchi}, M., {Mowlavi}, N., {et~al.}
  2023{\natexlab{b}}, \aap, 680, A36

\bibitem[{{Gavras} {et~al.}(2023){Gavras}, {Rimoldini}, {Nienartowicz}, {de
  Fombelle}, {Holl}, {{\'A}brah{\'a}m}, {Audard}, {Carnerero}, {Clementini},
  {De Ridder}, {Distefano}, {Garcia-Lario}, {Garofalo}, {K{\'o}sp{\'a}l},
  {Kruszy{\'n}ska}, {Kun}, {Lecoeur-Ta{\"\i}bi}, {Marton}, {Mazeh}, {Mowlavi},
  {Raiteri}, {Ripepi}, {Szabados}, {Zucker}, \& {Eyer}}]{GavrasEtal2023}
{Gavras}, P., {Rimoldini}, L., {Nienartowicz}, K., {et~al.} 2023, \aap, 674,
  A22

\bibitem[{{Gavras} {et~al.}(2022){Gavras}, {Rimoldini}, {Nienartowicz},
  {Jevardat de Fombelle}, {Holl}, {Abraham}, {Audard}, {Carnerero},
  {Clementini}, {De Ridder}, {Distefano}, {Garcia-Lario}, {Garofalo}, {Kospal},
  {Kruszynska}, {Kun}, {Lecoeur-Taibi}, {Marton}, {Mazeh}, {Mowlavi},
  {Raiteri}, {Ripepi}, {Szabados}, {Zucker}, \& {Eyer}}]{GavrasEtalCat2022}
{Gavras}, P., {Rimoldini}, L., {Nienartowicz}, K., {et~al.} 2022, VizieR Online
  Data Catalog, J/A+A/674/A22

\bibitem[{{Gomel} {et~al.}(2023){Gomel}, {Mazeh}, {Faigler}, {Bashi}, {Eyer},
  {Rimoldini}, {Audard}, {Mowlavi}, {Holl}, {Jevardat}, {Nienartowicz},
  {Lecoeur}, \& {Wyrzykowski}}]{GomelEtal2023}
{Gomel}, R., {Mazeh}, T., {Faigler}, S., {et~al.} 2023, \aap, 674, A19

\bibitem[{Hess(1912)}]{Hess1912}
Hess, V.~F. 1912, Phys. Z., 13, 1084

\bibitem[{{Hodgkin} {et~al.}(2021){Hodgkin}, {Harrison}, {Breedt}, {Wevers},
  {Rixon}, {Delgado}, {Yoldas}, {Kostrzewa-Rutkowska}, {Wyrzykowski}, {van
  Leeuwen}, {Blagorodnova}, {Campbell}, {Eappachen}, {Fraser}, {Ihanec},
  {Koposov}, {Kruszy{\'n}ska}, {Marton}, {Rybicki}, {Brown}, {Burgess},
  {Busso}, {Cowell}, {De Angeli}, {Diener}, {Evans}, {Gilmore}, {Holland},
  {Jonker}, {van Leeuwen}, {Mignard}, {Osborne}, {Portell}, {Prusti},
  {Richards}, {Riello}, {Seabroke}, {Walton}, {{\'A}brah{\'a}m}, {Altavilla},
  {Baker}, {Bastian}, {O'Brien}, {de Bruijne}, {Butterley}, {Carrasco},
  {Casta{\~n}eda}, {Clark}, {Clementini}, {Copperwheat}, {Cropper},
  {Damljanovic}, {Davidson}, {Davis}, {Dennefeld}, {Dhillon}, {Dolding},
  {Dominik}, {Esquej}, {Eyer}, {Fabricius}, {Fridman}, {Froebrich}, {Garralda},
  {Gomboc}, {Gonz{\'a}lez-Vidal}, {Guerra}, {Hambly}, {Hardy}, {Holl},
  {Hourihane}, {Japelj}, {Kann}, {Kiss}, {Knigge}, {Kolb}, {Komossa},
  {K{\'o}sp{\'a}l}, {Kov{\'a}cs}, {Kun}, {Leto}, {Lewis}, {Littlefair},
  {Mahabal}, {Mundell}, {Nagy}, {Padeletti}, {Palaversa}, {Pigulski},
  {Pretorius}, {van Reeven}, {Ribeiro}, {Roelens}, {Rowell}, {Schartel},
  {Scholz}, {Schwope}, {Sip{\H{o}}cz}, {Smartt}, {Smith}, {Serraller},
  {Steeghs}, {Sullivan}, {Szabados}, {Szegedi-Elek}, {Tisserand}, {Tomasella},
  {van Velzen}, {Whitelock}, {Wilson}, \& {Young}}]{HodgkinEtal2021}
{Hodgkin}, S.~T., {Harrison}, D.~L., {Breedt}, E., {et~al.} 2021, \aap, 652,
  A76

\bibitem[{{Holl} {et~al.}(2018){Holl}, {Audard}, {Nienartowicz}, {Jevardat de
  Fombelle}, {Marchal}, {Mowlavi}, {Clementini}, {De Ridder}, {Evans}, {Guy},
  {Lanzafame}, {Lebzelter}, {Rimoldini}, {Roelens}, {Zucker}, {Distefano},
  {Garofalo}, {Lecoeur-Ta{\"\i}bi}, {Lopez}, {Molinaro}, {Muraveva}, {Panahi},
  {Regibo}, {Ripepi}, {Sarro}, {Aerts}, {Anderson}, {Charnas}, {Barblan},
  {Blanco-Cuaresma}, {Busso}, {Cuypers}, {De Angeli}, {Glass}, {Grenon},
  {Juh{\'a}sz}, {Kochoska}, {Koubsky}, {Lanza}, {Leccia}, {Lorenz}, {Marconi},
  {Marschalk{\'o}}, {Mazeh}, {Messina}, {Mignard}, {Moitinho}, {Moln{\'a}r},
  {Morgenthaler}, {Musella}, {Ordenovic}, {Ord{\'o}{\~n}ez}, {Pagano},
  {Palaversa}, {Pawlak}, {Plachy}, {Pr{\v{s}}a}, {Riello}, {S{\"u}veges},
  {Szabados}, {Szegedi-Elek}, {Votruba}, \& {Eyer}}]{HollEtal2018}
{Holl}, B., {Audard}, M., {Nienartowicz}, K., {et~al.} 2018, \aap, 618, A30

\bibitem[{{Holl} {et~al.}(2023{\natexlab{a}}){Holl}, {Fabricius}, {Portell},
  {Lindegren}, {Panuzzo}, {Bernet}, {Casta{\~n}eda}, {Jevardat de Fombelle},
  {Audard}, {Ducourant}, {Harrison}, {Evans}, {Busso}, {Sozzetti}, {Gosset},
  {Arenou}, {De Angeli}, {Riello}, {Eyer}, {Rimoldini}, {Gavras}, {Mowlavi},
  {Nienartowicz}, {Lecoeur-Ta{\"\i}bi}, {Garc{\'\i}a-Lario}, \&
  {Pourbaix}}]{HollEtal2023}
{Holl}, B., {Fabricius}, C., {Portell}, J., {et~al.} 2023{\natexlab{a}}, \aap,
  674, A25

\bibitem[{{Holl} {et~al.}(2023{\natexlab{b}}){Holl}, {Sozzetti}, {Sahlmann},
  {Giacobbe}, {S{\'e}gransan}, {Unger}, {Delisle}, {Barbato}, {Lattanzi},
  {Morbidelli}, \& {Sosnowska}}]{HollEtal2023a}
{Holl}, B., {Sozzetti}, A., {Sahlmann}, J., {et~al.} 2023{\natexlab{b}}, \aap,
  674, A10

\bibitem[{{Ivezi{\'c}} {et~al.}(2019){Ivezi{\'c}}, {Kahn}, {Tyson}, {Abel},
  {Acosta}, {Allsman}, {Alonso}, {AlSayyad}, {Anderson}, {Andrew}, {Angel},
  {Angeli}, {Ansari}, {Antilogus}, {Araujo}, {Armstrong}, {Arndt}, {Astier},
  {Aubourg}, {Auza}, {Axelrod}, {Bard}, {Barr}, {Barrau}, {Bartlett}, {Bauer},
  {Bauman}, {Baumont}, {Bechtol}, {Bechtol}, {Becker}, {Becla}, {Beldica},
  {Bellavia}, {Bianco}, {Biswas}, {Blanc}, {Blazek}, {Blandford}, {Bloom},
  {Bogart}, {Bond}, {Booth}, {Borgland}, {Borne}, {Bosch}, {Boutigny},
  {Brackett}, {Bradshaw}, {Brandt}, {Brown}, {Bullock}, {Burchat}, {Burke},
  {Cagnoli}, {Calabrese}, {Callahan}, {Callen}, {Carlin}, {Carlson},
  {Chandrasekharan}, {Charles-Emerson}, {Chesley}, {Cheu}, {Chiang}, {Chiang},
  {Chirino}, {Chow}, {Ciardi}, {Claver}, {Cohen-Tanugi}, {Cockrum}, {Coles},
  {Connolly}, {Cook}, {Cooray}, {Covey}, {Cribbs}, {Cui}, {Cutri}, {Daly},
  {Daniel}, {Daruich}, {Daubard}, {Daues}, {Dawson}, {Delgado}, {Dellapenna},
  {de Peyster}, {de Val-Borro}, {Digel}, {Doherty}, {Dubois},
  {Dubois-Felsmann}, {Durech}, {Economou}, {Eifler}, {Eracleous}, {Emmons},
  {Fausti Neto}, {Ferguson}, {Figueroa}, {Fisher-Levine}, {Focke}, {Foss},
  {Frank}, {Freemon}, {Gangler}, {Gawiser}, {Geary}, {Gee}, {Geha}, {Gessner},
  {Gibson}, {Gilmore}, {Glanzman}, {Glick}, {Goldina}, {Goldstein}, {Goodenow},
  {Graham}, {Gressler}, {Gris}, {Guy}, {Guyonnet}, {Haller}, {Harris},
  {Hascall}, {Haupt}, {Hernandez}, {Herrmann}, {Hileman}, {Hoblitt}, {Hodgson},
  {Hogan}, {Howard}, {Huang}, {Huffer}, {Ingraham}, {Innes}, {Jacoby}, {Jain},
  {Jammes}, {Jee}, {Jenness}, {Jernigan}, {Jevremovi{\'c}}, {Johns}, {Johnson},
  {Johnson}, {Jones}, {Juramy-Gilles}, {Juri{\'c}}, {Kalirai}, {Kallivayalil},
  {Kalmbach}, {Kantor}, {Karst}, {Kasliwal}, {Kelly}, {Kessler}, {Kinnison},
  {Kirkby}, {Knox}, {Kotov}, {Krabbendam}, {Krughoff}, {Kub{\'a}nek},
  {Kuczewski}, {Kulkarni}, {Ku}, {Kurita}, {Lage}, {Lambert}, {Lange},
  {Langton}, {Le Guillou}, {Levine}, {Liang}, {Lim}, {Lintott}, {Long},
  {Lopez}, {Lotz}, {Lupton}, {Lust}, {MacArthur}, {Mahabal}, {Mandelbaum},
  {Markiewicz}, {Marsh}, {Marshall}, {Marshall}, {May}, {McKercher}, {McQueen},
  {Meyers}, {Migliore}, {Miller}, {Mills}, {Miraval}, {Moeyens}, {Moolekamp},
  {Monet}, {Moniez}, {Monkewitz}, {Montgomery}, {Morrison}, {Mueller},
  {Muller}, {Mu{\~n}oz Arancibia}, {Neill}, {Newbry}, {Nief}, {Nomerotski},
  {Nordby}, {O'Connor}, {Oliver}, {Olivier}, {Olsen}, {O'Mullane}, {Ortiz},
  {Osier}, {Owen}, {Pain}, {Palecek}, {Parejko}, {Parsons}, {Pease},
  {Peterson}, {Peterson}, {Petravick}, {Libby Petrick}, {Petry},
  {Pierfederici}, {Pietrowicz}, {Pike}, {Pinto}, {Plante}, {Plate}, {Plutchak},
  {Price}, {Prouza}, {Radeka}, {Rajagopal}, {Rasmussen}, {Regnault}, {Reil},
  {Reiss}, {Reuter}, {Ridgway}, {Riot}, {Ritz}, {Robinson}, {Roby}, {Roodman},
  {Rosing}, {Roucelle}, {Rumore}, {Russo}, {Saha}, {Sassolas}, {Schalk},
  {Schellart}, {Schindler}, {Schmidt}, {Schneider}, {Schneider}, {Schoening},
  {Schumacher}, {Schwamb}, {Sebag}, {Selvy}, {Sembroski}, {Seppala}, {Serio},
  {Serrano}, {Shaw}, {Shipsey}, {Sick}, {Silvestri}, {Slater}, {Smith},
  {Smith}, {Sobhani}, {Soldahl}, {Storrie-Lombardi}, {Stover}, {Strauss},
  {Street}, {Stubbs}, {Sullivan}, {Sweeney}, {Swinbank}, {Szalay}, {Takacs},
  {Tether}, {Thaler}, {Thayer}, {Thomas}, {Thornton}, {Thukral}, {Tice},
  {Trilling}, {Turri}, {Van Berg}, {Vanden Berk}, {Vetter}, {Virieux},
  {Vucina}, {Wahl}, {Walkowicz}, {Walsh}, {Walter}, {Wang}, {Wang}, {Warner},
  {Wiecha}, {Willman}, {Winters}, {Wittman}, {Wolff}, {Wood-Vasey}, {Wu},
  {Xin}, {Yoachim}, \& {Zhan}}]{IvezicEtal2019}
{Ivezi{\'c}}, {\v{Z}}., {Kahn}, S.~M., {Tyson}, J.~A., {et~al.} 2019, \apj,
  873, 111

\bibitem[{{Lebzelter} {et~al.}(2023){Lebzelter}, {Mowlavi}, {Lecoeur-Taibi},
  {Trabucchi}, {Audard}, {Garc{\'\i}a-Lario}, {Gavras}, {Holl}, {Jevardat de
  Fombelle}, {Nienartowicz}, {Rimoldini}, \& {Eyer}}]{LebzelterEtal2023}
{Lebzelter}, T., {Mowlavi}, N., {Lecoeur-Taibi}, I., {et~al.} 2023, \aap, 674,
  A15

\bibitem[{{Mainieri} {et~al.}(2024){Mainieri}, {Anderson}, {Brinchmann},
  {Cimatti}, {Ellis}, {Hill}, {Kneib}, {McLeod}, {Opitom}, {Roth},
  {Sanchez-Saez}, {Smilljanic}, {Tolstoy}, {Bacon}, {Randich}, {Adamo},
  {Annibali}, {Arevalo}, {Audard}, {Barsanti}, {Battaglia}, {Bayo Aran},
  {Belfiore}, {Bellazzini}, {Bellini}, {Beltran}, {Berni}, {Bianchi}, {Biazzo},
  {Bisero}, {Bisogni}, {Bland-Hawthorn}, {Blondin}, {Bodensteiner}, {Boffin},
  {Bonito}, {Bono}, {Bouche}, {Bowman}, {Braga}, {Bragaglia}, {Branchesi},
  {Brucalassi}, {Bryant}, {Bryson}, {Busa}, {Camera}, {Carbone}, {Casali},
  {Casali}, {Casasola}, {Castro}, {Catelan}, {Cavallo}, {Chiappini}, {Cioni},
  {Colless}, {Colzi}, {Contarini}, {Couch}, {D'Ammando}, {d'Assignies D.},
  {D'Orazi}, {da Silva}, {Dainotti}, {Damiani}, {Danielski}, {De Cia}, {de
  Jong}, {Dhawan}, {Dierickx}, {Driver}, {Dupletsa}, {Escoffier}, {Escorza},
  {Fabrizio}, {Fiorentino}, {Fontana}, {Fontani}, {Forero Sanchez}, {Franois},
  {Galindo-Guil}, {Gallazzi}, {Galli}, {Garcia}, {Garcia-Rojas}, {Garilli},
  {Grand}, {Guarcello}, {Hazra}, {Helmi}, {Herrero}, {Iglesias}, {Ilic},
  {Irsic}, {Ivanov}, {Izzo}, {Jablonka}, {Joachimi}, {Kakkad}, {Kamann},
  {Koposov}, {Kordopatis}, {Kovacevic}, {Kraljic}, {Kuncarayakti}, {Kwon}, {La
  Forgia}, {Lahav}, {Laigle}, {Lazzarin}, {Leaman}, {Leclercq}, {Lee}, {Lee},
  {Lehnert}, {Lira}, {Loffredo}, {Lucatello}, {Magrini}, {Maguire}, {Mahler},
  {Zahra Majidi}, {Malavasi}, {Mannucci}, {Marconi}, {Martin}, {Marulli},
  {Massari}, {Matsuno}, {Mattheee}, {McGee}, {Merc}, {Merle}, {Miglio},
  {Migliorini}, {Minchev}, {Minniti}, {Miret-Roig}, {Monreal Ibero}, {Montano},
  {Montet}, {Moresco}, {Moretti}, {Moscardini}, {Moya}, {Mueller},
  {Nanayakkara}, {Nicholl}, {Nordlander}, {Onori}, {Padovani}, {Pala}, {Panda},
  {Pandey-Pommier}, {Pasquini}, {Pawlak}, {Pessi}, {Pisani}, {Popovic},
  {Prisinzano}, {Raddi}, {Rainer}, {Rebassa-Mansergas}, {Richard}, {Rigault},
  {Rocher}, {Romano}, {Rosati}, {Sacco}, {Sanchez-Janssen}, {Sander},
  {Sanders}, {Sargent}, {Sarpa}, {Schimd}, {Schipani}, {Sefusatti}, {Smith},
  {Spina}, {Steinmetz}, {Tacchella}, {Tautvaisiene}, {Theissen}, {Thomas},
  {Ting}, {Travouillon}, {Tresse}, {Trivedi}, {Tsantaki}, {Tsedrik}, {Urrutia},
  {Valenti}, {Van der Swaelmen}, {Van Eck}, {Verdiani}, {Verdier}, {Vergani},
  {Verhamme}, {Vernet}, {Verza}, {Viel}, {Vielzeuf}, {Vietri}, {Vink},
  {Viscasillas Vazquez}, {Wang}, {Weilbacher}, {Wendt}, {Wright}, {Ye},
  {Yeche}, {Yu}, {Zafar}, {Zibetti}, {Ziegler}, \&
  {Zinchenko}}]{MainieriEtal2024}
{Mainieri}, V., {Anderson}, R.~I., {Brinchmann}, J., {et~al.} 2024, arXiv
  e-prints, arXiv:2403.05398

\bibitem[{{Marton} {et~al.}(2023){Marton}, {{\'A}brah{\'a}m}, {Rimoldini},
  {Audard}, {Kun}, {Nagy}, {K{\'o}sp{\'a}l}, {Szabados}, {Holl}, {Gavras},
  {Mowlavi}, {Nienartowicz}, {de Fombelle}, {Lecoeur-Ta{\"\i}bi}, {Karbevska},
  {Lario}, \& {Eyer}}]{MartonEtal2023}
{Marton}, G., {{\'A}brah{\'a}m}, P., {Rimoldini}, L., {et~al.} 2023, \aap, 674,
  A21

\bibitem[{{Mer{\'\i}n} {et~al.}(2015){Mer{\'\i}n}, {Salgado}, {Giordano},
  {Baines}, {Sarmiento}, {L{\'o}pez Mart{\'\i}}, {Racero}, {Guti{\'e}rrez},
  {Pollock}, {Rosa}, {Castellanos}, {Gonz{\'a}lez}, {Le{\'o}n}, {Ortiz de
  Landaluce}, {de Teodoro}, {Nieto}, {Lennon}, {Arviset}, {de Marchi}, \&
  {O'Mullane}}]{MerinEtal2015}
{Mer{\'\i}n}, B., {Salgado}, J., {Giordano}, F., {et~al.} 2015, arXiv e-prints,
  arXiv:1512.00842

\bibitem[{{Mowlavi} {et~al.}(2023){Mowlavi}, {Holl}, {Lecoeur-Ta{\"\i}bi},
  {Barblan}, {Kochoska}, {Pr{\v{s}}a}, {Mazeh}, {Rimoldini}, {Gavras},
  {Audard}, {Jevardat de Fombelle}, {Nienartowicz}, {Garc{\'\i}a-Lario}, \&
  {Eyer}}]{MowlaviEtal2023}
{Mowlavi}, N., {Holl}, B., {Lecoeur-Ta{\"\i}bi}, I., {et~al.} 2023, \aap, 674,
  A16

\bibitem[{{Panahi} {et~al.}(2022{\natexlab{a}}){Panahi}, {Mazeh}, {Zucker},
  {Latham}, {Collins}, {Rimoldini}, {Evans}, \& {Eyer}}]{PanahiEtal2022b}
{Panahi}, A., {Mazeh}, T., {Zucker}, S., {et~al.} 2022{\natexlab{a}}, \aap,
  667, A14

\bibitem[{{Panahi} {et~al.}(2022{\natexlab{b}}){Panahi}, {Zucker},
  {Clementini}, {Audard}, {Binnenfeld}, {Cusano}, {Evans}, {Gomel}, {Holl},
  {Ilyin}, {de Fombelle}, {Mazeh}, {Mowlavi}, {Nienartowicz}, {Rimoldini},
  {Shahaf}, \& {Eyer}}]{PanahiEtal2022a}
{Panahi}, A., {Zucker}, S., {Clementini}, G., {et~al.} 2022{\natexlab{b}},
  \aap, 663, A101

\bibitem[{{Pogson}(1856)}]{Pogson1856}
{Pogson}, N. 1856, \mnras, 17, 12

\bibitem[{{Pourbaix} {et~al.}(2022){Pourbaix}, {Arenou}, {Gavras}, {Gosset},
  {Halbwachs}, {Siopis}, {Sozzetti}, {Bauchet}, {Damerdji}, {Delchambre},
  {Delisle}, {Giacobbe}, {Holl}, {Jorissen}, {Lattanzi}, {Leclerc}, {Morel},
  {Sadowski}, {Sahlmann}, \& {Segransan}}]{PourbaixEtal2022}
{Pourbaix}, D., {Arenou}, F., {Gavras}, P., {et~al.} 2022, {Gaia DR3
  documentation Chapter 7: Non-single stars}, Gaia DR3 documentation, European
  Space Agency; Gaia Data Processing and Analysis Consortium. Online at <A
  href=``https://gea.esac.esa.int/archive/documentation/GDR3/index.html''>https://gea.esac.esa.int/archive/documentation/GDR3/index.html</A>,
  id. 7

\bibitem[{{Rauer} {et~al.}(2014){Rauer}, {Catala}, {Aerts}, {Appourchaux},
  {Benz}, {Brandeker}, {Christensen-Dalsgaard}, {Deleuil}, {Gizon}, {Goupil},
  {G{\"u}del}, {Janot-Pacheco}, {Mas-Hesse}, {Pagano}, {Piotto}, {Pollacco},
  {Santos}, {Smith}, {Su{\'a}rez}, {Szab{\'o}}, {Udry}, {Adibekyan}, {Alibert},
  {Almenara}, {Amaro-Seoane}, {Eiff}, {Asplund}, {Antonello}, {Barnes},
  {Baudin}, {Belkacem}, {Bergemann}, {Bihain}, {Birch}, {Bonfils}, {Boisse},
  {Bonomo}, {Borsa}, {Brand{\~a}o}, {Brocato}, {Brun}, {Burleigh}, {Burston},
  {Cabrera}, {Cassisi}, {Chaplin}, {Charpinet}, {Chiappini}, {Church},
  {Csizmadia}, {Cunha}, {Damasso}, {Davies}, {Deeg}, {D{\'\i}az}, {Dreizler},
  {Dreyer}, {Eggenberger}, {Ehrenreich}, {Eigm{\"u}ller}, {Erikson}, {Farmer},
  {Feltzing}, {de Oliveira Fialho}, {Figueira}, {Forveille}, {Fridlund},
  {Garc{\'\i}a}, {Giommi}, {Giuffrida}, {Godolt}, {Gomes da Silva}, {Granzer},
  {Grenfell}, {Grotsch-Noels}, {G{\"u}nther}, {Haswell}, {Hatzes},
  {H{\'e}brard}, {Hekker}, {Helled}, {Heng}, {Jenkins}, {Johansen},
  {Khodachenko}, {Kislyakova}, {Kley}, {Kolb}, {Krivova}, {Kupka}, {Lammer},
  {Lanza}, {Lebreton}, {Magrin}, {Marcos-Arenal}, {Marrese}, {Marques},
  {Martins}, {Mathis}, {Mathur}, {Messina}, {Miglio}, {Montalban}, {Montalto},
  {Monteiro}, {Moradi}, {Moravveji}, {Mordasini}, {Morel}, {Mortier},
  {Nascimbeni}, {Nelson}, {Nielsen}, {Noack}, {Norton}, {Ofir}, {Oshagh},
  {Ouazzani}, {P{\'a}pics}, {Parro}, {Petit}, {Plez}, {Poretti}, {Quirrenbach},
  {Ragazzoni}, {Raimondo}, {Rainer}, {Reese}, {Redmer}, {Reffert},
  {Rojas-Ayala}, {Roxburgh}, {Salmon}, {Santerne}, {Schneider}, {Schou},
  {Schuh}, {Schunker}, {Silva-Valio}, {Silvotti}, {Skillen}, {Snellen}, {Sohl},
  {Sousa}, {Sozzetti}, {Stello}, {Strassmeier}, {{\v{S}}vanda}, {Szab{\'o}},
  {Tkachenko}, {Valencia}, {Van Grootel}, {Vauclair}, {Ventura}, {Wagner},
  {Walton}, {Weingrill}, {Werner}, {Wheatley}, \& {Zwintz}}]{RauerEtal2014}
{Rauer}, H., {Catala}, C., {Aerts}, C., {et~al.} 2014, Experimental Astronomy,
  38, 249

\bibitem[{{Ricker} {et~al.}(2015){Ricker}, {Winn}, {Vanderspek}, {Latham},
  {Bakos}, {Bean}, {Berta-Thompson}, {Brown}, {Buchhave}, {Butler}, {Butler},
  {Chaplin}, {Charbonneau}, {Christensen-Dalsgaard}, {Clampin}, {Deming},
  {Doty}, {De Lee}, {Dressing}, {Dunham}, {Endl}, {Fressin}, {Ge}, {Henning},
  {Holman}, {Howard}, {Ida}, {Jenkins}, {Jernigan}, {Johnson}, {Kaltenegger},
  {Kawai}, {Kjeldsen}, {Laughlin}, {Levine}, {Lin}, {Lissauer}, {MacQueen},
  {Marcy}, {McCullough}, {Morton}, {Narita}, {Paegert}, {Palle}, {Pepe},
  {Pepper}, {Quirrenbach}, {Rinehart}, {Sasselov}, {Sato}, {Seager},
  {Sozzetti}, {Stassun}, {Sullivan}, {Szentgyorgyi}, {Torres}, {Udry}, \&
  {Villasenor}}]{RickerEtal2015}
{Ricker}, G.~R., {Winn}, J.~N., {Vanderspek}, R., {et~al.} 2015, Journal of
  Astronomical Telescopes, Instruments, and Systems, 1, 014003

\bibitem[{{Riello} {et~al.}(2021){Riello}, {De Angeli}, {Evans}, {Montegriffo},
  {Carrasco}, {Busso}, {Palaversa}, {Burgess}, {Diener}, {Davidson}, {Rowell},
  {Fabricius}, {Jordi}, {Bellazzini}, {Pancino}, {Harrison}, {Cacciari}, {van
  Leeuwen}, {Hambly}, {Hodgkin}, {Osborne}, {Altavilla}, {Barstow}, {Brown},
  {Castellani}, {Cowell}, {De Luise}, {Gilmore}, {Giuffrida}, {Hidalgo},
  {Holland}, {Marinoni}, {Pagani}, {Piersimoni}, {Pulone}, {Ragaini}, {Rainer},
  {Richards}, {Sanna}, {Walton}, {Weiler}, \& {Yoldas}}]{RielloEtal2021}
{Riello}, M., {De Angeli}, F., {Evans}, D.~W., {et~al.} 2021, \aap, 649, A3

\bibitem[{{Rimoldini} {et~al.}(2023){Rimoldini}, {Holl}, {Gavras}, {Audard},
  {De Ridder}, {Mowlavi}, {Nienartowicz}, {Jevardat de Fombelle},
  {Lecoeur-Ta{\"\i}bi}, {Karbevska}, {Evans}, {{\'A}brah{\'a}m}, {Carnerero},
  {Clementini}, {Distefano}, {Garofalo}, {Garc{\'\i}a-Lario}, {Gomel},
  {Klioner}, {Kruszy{\'n}ska}, {Lanzafame}, {Lebzelter}, {Marton}, {Mazeh},
  {Molinaro}, {Panahi}, {Raiteri}, {Ripepi}, {Szabados}, {Teyssier},
  {Trabucchi}, {Wyrzykowski}, {Zucker}, \& {Eyer}}]{RimoldiniEtal2023}
{Rimoldini}, L., {Holl}, B., {Gavras}, P., {et~al.} 2023, \aap, 674, A14

\bibitem[{{Ripepi} {et~al.}(2023){Ripepi}, {Clementini}, {Molinaro}, {Leccia},
  {Plachy}, {Moln{\'a}r}, {Rimoldini}, {Musella}, {Marconi}, {Garofalo},
  {Audard}, {Holl}, {Evans}, {Jevardat de Fombelle}, {Lecoeur-Taibi},
  {Marchal}, {Mowlavi}, {Muraveva}, {Nienartowicz}, {Sartoretti}, {Szabados},
  \& {Eyer}}]{RipepiEtal2023}
{Ripepi}, V., {Clementini}, G., {Molinaro}, R., {et~al.} 2023, \aap, 674, A17

\bibitem[{Sahu {et~al.}(2022)Sahu, Anderson, Casertano, Bond, Udalski, Dominik,
  Calamida, Bellini, Brown, Rejkuba, Bajaj, Kains, Ferguson, Fryer, Yock,
  Mróz, Kozłowski, Pietrukowicz, Poleski, Skowron, Soszyński, Szymański,
  Ulaczyk, Łukasz Wyrzykowski, Collaboration), Barry, Bennett, Bond, Hirao,
  Silva, Kondo, Koshimoto, Ranc, Rattenbury, Sumi, Suzuki, Tristram, Vandorou,
  Collaboration), Beaulieu, Marquette, Cole, Fouqué, Hill, Dieters, Coutures,
  Dominis-Prester, Bennett, Bachelet, Menzies, Albrow, Pollard, Collaboration),
  Gould, Yee, Allen, Almeida, Christie, Drummond, Gal-Yam, Gorbikov, Jablonski,
  Lee, Maoz, Manulis, McCormick, Natusch, Pogge, Shvartzvald, Collaboration),
  Jørgensen, Alsubai, Andersen, Bozza, Novati, Burgdorf, Hinse, Hundertmark,
  Husser, Kerins, Longa-Peña, Mancini, Penny, Rahvar, Ricci, Sajadian,
  Skottfelt, Snodgrass, Southworth, Tregloan-Reed, Wambsganss, Wertz,
  Consortium), Tsapras, Street, Bramich, Horne, Steele, \&
  Collaboration)}]{SahuEtal2022}
Sahu, K.~C., Anderson, J., Casertano, S., {et~al.} 2022, The Astrophysical
  Journal, 933, 83

\bibitem[{{Schatzman}(1963)}]{Schatzman1963}
{Schatzman}, E. 1963, {Astrophysique} (Masson et Cie)

\bibitem[{{Sozzetti} {et~al.}(2023){Sozzetti}, {Pinamonti}, {Damasso},
  {Desidera}, {Biazzo}, {Bonomo}, {Nardiello}, {Gratton}, {Lanza}, {Malavolta},
  {Giacobbe}, {Affer}, {Bignamini}, {Borsa}, {Boschin}, {Brogi}, {Cabona},
  {Claudi}, {Covino}, {Di Fabrizio}, {Ghedina}, {Harutyunyan}, {Knapic},
  {Maldonado}, {Maggio}, {Mancini}, {Mantovan}, {Marzari}, {Messina}, {Micela},
  {Molinari}, {Montalto}, {Naponiello}, {Pagano}, {Pedani}, {Piotto},
  {Poretti}, {Scandariato}, {Silvotti}, \& {Turrini}}]{SozzezziEtal2023}
{Sozzetti}, A., {Pinamonti}, M., {Damasso}, M., {et~al.} 2023, \aap, 677, L15

\bibitem[{{S{\"u}veges} {et~al.}(2012){S{\"u}veges}, {Sesar}, {V{\'a}radi},
  {Mowlavi}, {Becker}, {Ivezi{\'c}}, {Beck}, {Nienartowicz}, {Rimoldini},
  {Dubath}, {Bartholdi}, \& {Eyer}}]{SuvegesEtal2012}
{S{\"u}veges}, M., {Sesar}, B., {V{\'a}radi}, M., {et~al.} 2012, \mnras, 424,
  2528

\bibitem[{{Taylor}(2005)}]{Taylor2005}
{Taylor}, M.~B. 2005, in Astronomical Society of the Pacific Conference Series,
  Vol. 347, Astronomical Data Analysis Software and Systems XIV, ed.
  P.~{Shopbell}, M.~{Britton}, \& R.~{Ebert}, 29

\bibitem[{{Udalski}(2003)}]{OGLE2003}
{Udalski}, A. 2003, \actaa, 53, 291

\bibitem[{{Wenger} {et~al.}(2000){Wenger}, {Ochsenbein}, {Egret}, {Dubois},
  {Bonnarel}, {Borde}, {Genova}, {Jasniewicz}, {Lalo{\"e}}, {Lesteven}, \&
  {Monier}}]{Simbad2000}
{Wenger}, M., {Ochsenbein}, F., {Egret}, D., {et~al.} 2000, \aaps, 143, 9

\bibitem[{{Wu} {et~al.}(2023){Wu}, {Dong}, {Yi}, {Liu}, {El-Badry}, {Gould},
  {Christie}, {de Almeida}, {Monard}, {McCormick}, {Chen}, {Huang}, {Liu},
  {Merand}, {Mroz}, {Shangguan}, {Udalski}, {Woillez}, \& {Zhang}}]{WuEtal2023}
{Wu}, Z., {Dong}, S., {Yi}, T., {et~al.} 2023, arXiv e-prints, arXiv:2309.03944

\bibitem[{{Wyrzykowski} {et~al.}(2023){Wyrzykowski}, {Kruszy{\'n}ska},
  {Rybicki}, {Holl}, {Lec{\oe}ur-Ta{\"\i}bi}, {Mowlavi}, {Nienartowicz},
  {Jevardat de Fombelle}, {Rimoldini}, {Audard}, {Garcia-Lario}, {Gavras},
  {Evans}, {Hodgkin}, \& {Eyer}}]{WyrzykowskiEtal2023}
{Wyrzykowski}, {\L}., {Kruszy{\'n}ska}, K., {Rybicki}, K.~A., {et~al.} 2023,
  \aap, 674, A23

\end{thebibliography}
\end{document}